\documentclass[a4paper,11pt,nohyper]{article}
\pdfoutput=1
\usepackage{jheppub}

\usepackage[T1]{fontenc}
\usepackage[mathletters]{ucs} 
\usepackage[utf8x]{inputenc}
\usepackage{multirow,booktabs,array}
\usepackage{amssymb}
\usepackage{amsmath}
\usepackage{lmodern}
\usepackage{xcolor}
\usepackage{braket}
\usepackage{verbatim}
\usepackage{multirow}
\usepackage{graphics,graphicx}
\usepackage{psfrag}
\usepackage{empheq} 
\usepackage{mathtools}
\usepackage{cleveref}

\def\a{\alpha}
\def\b{\beta}

\def\q{\theta}                    

\let\a=\alpha \let\b=\beta   
  \let\q=\theta

\def\nn{\nonumber} \def\bd{\begin{document}} \def\ed{\end{document}}
\def\ds{\documentstyle} \let\fr=\frac \let\bl=\bigl \let\br=\bigr
\let\Br=\Bigr \let\Bl=\Bigl
\let\bm=\bibitem
\let\na=\nabla

\newcommand{\boxedeq}[2]{\begin{empheq}[box={\fboxsep=6pt\fbox}]{align}\label{#1}#2\end{empheq}}
\newcommand{\be}{\begin{equation}}
\newcommand{\ee}{\end{equation}}

\newcommand{\bea}{\begin{eqnarray}}
\newcommand{\eea}{\end{eqnarray}}

\definecolor{ao(english)}{rgb}{0.0, 0.5, 0.0}

\usepackage{amsmath,latexsym,amssymb,slashed}
\usepackage{mathrsfs}

\title{Cosmic branes and asymptotic structure}
 \author{F. Capone}
  \author{and M. Taylor}
 \affiliation{Mathematical Sciences and STAG Research Centre, University of Southampton\\Highfield, Southampton, SO17 1BJ, UK.} 
\bigskip
\emailAdd{f.capone@soton.ac.uk}
\emailAdd{ m.m.taylor@soton.ac.uk}

\abstract{
Superrotations of asymptotically flat spacetimes in four dimensions can be interpreted in terms of including cosmic strings within the phase space of allowed solutions. In this paper we explore the implications of the inclusion of cosmic branes on the asymptotic structure of vacuum spacetimes in dimension $d > 4$. We first show that only cosmic $(d-3)$-branes are Riemann flat in the neighbourhood of the brane, and therefore only branes of such dimension passing through the celestial sphere can respect asymptotic local flatness. We derive the asymptotically locally flat boundary conditions associated with including cosmic branes in the phase space of solutions. We find the asymptotic expansion of vacuum spacetimes in $d=5$ with such boundary conditions; the expansion is polyhomogenous, with logarithmic terms arising at subleading orders in the expansion. The asymptotically locally flat boundary conditions identified here are associated with an extended asymptotic symmetry group, which may be relevant to soft scattering theorems and memory effects.  
}

\preprint{}

\begin{document}
	\flushbottom
	\maketitle
	
	
	
	
	\section{Introduction}

Twenty years after the original AdS/CFT examples of holography were discovered by Maldacena \cite{Maldacena:1997re}, a broad landscape of gauge/gravity 
dualities has been uncovered, spanning non-conformal and non-relativistic quantum field theories. Almost all examples for which a detailed holographic 
dictionary has been constructed share a common feature: the quantum field theories are associated with timelike (conformal) boundaries of the bulk spacetimes. 

The formulation of holography for spacetimes whose boundaries are not timelike is conceptually challenging, but this is clearly important physically to describe both flat spacetimes and cosmologies. There has been considerable work on the latter, most of which makes use of analytic continuations of AdS/CFT and related dualities, see the original dS/CFT correspondence \cite{Strominger:2001pn} and precision holography for cosmology \cite{McFadden:2009fg}. 

Holography for asymptotically flat spacetimes is even more challenging. The conformal boundary consists of both future and past null infinity; spacelike infinity and future and past timelike infinity; it is thus a priori far from clear how one could associate a Lorentzian quantum field theory in one less dimension with the bulk spacetime. Moreover, it was shown many years ago in \cite{Skenderis:2000in} that the asymptotic structure near spacelike infinity cannot be associated with a local (Riemannian signature) theory associated with spacelike infinity. In \cite{Witten:2001kn} it was suggested that the analogue of holography for flat spacetimes might simply be S matrix relations. In other holographic dualities, the asymptotic symmetry group underpins the duality, and accordingly attention turned to studying the implications of the 
Bondi-Metzner-Sachs (BMS) asymptotic symmetry group of flat spacetimes \cite{Bondi1962,Sachs1962,Sachs1962a}. 

A surge of interest in this topic followed the works of \cite{Barnich2010a,Barnich2010, Barnich2011,Barnich2011a} on superrotations (see also \cite{Banks2003}). These works revisited the asymptotic symmetry group, arguing that in addition to the BMS symmetries, the symmetry group should include superrotation transformations that are meromorphic on the celestial sphere. Further discussions of the associated conserved charges can be found in \cite{FlanaganBMS,Flanagan2018}.
Subsequent works explored the relationship between asymptotic symmetries and soft scattering, see \cite{Cachazo2014,Campiglia2014,Strominger2014, Campiglia2015,He2015, Strominger2016, Pasterski2016,Distler2018} and the review \cite{StromingerLect2017}. 
It has also been proposed that BMS symmetries associated with black hole horizons are relevant to understanding the microstates of black holes, and hence to resolving the information loss paradox, see \cite{Hawking2017,Haco:2018ske}. 

There are still many open issues associated with asymptotically flat spacetimes in four dimensions. Various extensions of the BMS symmetry group have been proposed, with corresponding asymptotic boundary conditions on the spacetime. The boundary conditions and symmetry group proposed by \cite{Campiglia2014,Campiglia2015} does not preserve the Bondi-Penrose notions of asymptotic flatness, but the Ward identities associated with the symmetry group do indeed give the Cachazo-Strominger \cite{Cachazo2014} scattering theorems. An issue with superrotations is that they are defined as infinitesimal transformations. The authors of \cite{Compere2016} constructed vacua associated with finite superrotation transformations (of the type considered by \cite{Barnich2010a,Barnich2010, Barnich2011,Barnich2011a}) but the associated energies were ill-defined and did not seem to be bounded from below.  

In \cite{Strominger2017}, a physical interpretation of superrotations was given: it was argued that the poles in the meromorphic symmetry transformations are associated with cosmic strings piercing the celestial sphere. Indeed, meromorphic transformations on the celestial sphere were already discussed in the cosmic string literature, 
see \cite{Podolsky2000} for a study of how the collision and snapping of cosmic strings generates gravitational waves. A physical motivation for allowing general enough boundary conditions for the asymptotic symmetry group to include superrotations is hence to ensure that one includes in the phase space solutions such as Robinson-Trautman and their impulsive limits (i.e. snapping cosmic strings). The construction of such phase spaces was explored in \cite{Compere2018,Distler2018}; see also earlier work \cite{Ashtekar:1981ar}. 

The detailed relationship between superrotations and cosmic strings in four dimensions is not yet complete. In particular, in the derivations of \cite{Distler2018} it is assumed that the only singularities in the complex plane are associated with infinity; singularities at finite points are more subtle and are considered. However, from the work \cite{Podolsky2000}, it is only possible to use conformal transformations to move all singularities to infinity in the complex plane when there is only a single cosmic string; in the presence of multiple cosmic strings not all of the singularities can be sent to infinity in the complex plane. 

\bigskip

This paper is about boundary conditions and corresponding asymptotic symmetries for asymptotically (locally) flat spacetimes in dimensions higher than four. There is a long history of studying asymptotically flat spacetimes in dimensions higher than four, see for example \cite{Hollands2003,Hollands2004,Hollands2005,Tanabe:2009xb,Tanabe:2009va,Tanabe:2010rm,Tanabe2011,Tanabe2012}. This topic has been revisited recently in the context of relating asymptotic symmetries to soft scattering \cite{Kapec2015, Hollands2016,Delmastro2017}, while the role of extended symmetry groups in characterising soft hair for higher dimensional black holes was 
discussed in \cite{Shi2016}. Superrotations (and supertranslations) were extended to higher spin theories in \cite{Campoleoni2017}, and also considered higher dimensional asymptotically flat spacetimes.

A puzzling feature is that while soft scattering theorems exist in all such spacetime dimensions there seems to be no analogue of superrotations in dimensions higher than four. The original constructions of 
 \cite{Barnich2010a,Barnich2010, Barnich2011,Barnich2011a} were clearly specific to four dimensions and rely implicitly on the celestial sphere being two-dimensional: the superrotations are not analytic, but are meromorphic. For higher dimensional celestial spheres one cannot use complex analysis to classify allowable non-analytic symmetry transformations.  

Following the relation between cosmic strings and superrotations discussed in \cite{Strominger2017}, we use cosmic branes to define allowed boundary conditions for asymptotically (locally) flat spacetimes in $d > 4$. In $d =4$ cosmic strings are Riemann flat except at the location of the string; all higher dimensional defects are not Riemann flat in the vicinity of the string and indeed are not even asymptotically locally flat. In section \ref{sec:onetwo} we explore cosmic branes in $d > 4$, following the original approach of Vilenkin for cosmic strings \cite{Vilenkin:1981zs}. We point out that in $d$ dimensions only cosmic $(d-3)$-branes are Riemann flat in the vicinity of the brane. (Note that there are distributional curvature singularities at the location of the brane, as for cosmic strings in four dimensions.) 
Therefore the direct analogue of cosmic strings in four dimensions is cosmic $(d-3)$-branes in $d$ dimensions. Other types of cosmic branes are not locally Riemann flat near the brane; if such a brane pierces the celestial sphere, the geometry in the vicinity is not locally Riemann flat, and hence the resulting spacetime is not asymptotically locally flat. 

This observation is consistent with the fact that higher dimensional generalisations of metrics describing cosmic strings snapping have never been found. For example, in \cite{Podolsky:2006du} higher-dimensional generalisations of Robinson-Trautman spacetimes were constructed. There are no type $N$ spherical gravitational waves in this class and because of this there is no impulsive limit; nor did \cite{Podolsky:2006du} find  an analogue of the four-dimensional $C$ metric. It would be interesting to explore whether the class of solutions constructed in \cite{Podolsky:2006du} could accommodate cosmic branes breaking: this seems quite likely, as the transverse spatial part of the metric is an arbitrary Riemann Einstein space, just as we find for our asymptotic solutions described below. 

\bigskip

In section \ref{sec:Three} we analyse the asymptotic behaviour of cosmic branes, focussing for definiteness on the example of cosmic membranes in five dimensions. Following analogous discussions to those in \cite{Podolsky2000,Strominger2017}, we consider processes in which cosmic branes can snap, and infer the associated boundary conditions. In four dimensions, cosmic string snapping is consistent with asymptotically flat boundary conditions. In $d > 4$, the inclusion of snapping cosmic branes in the phase space requires asymptotically locally flat boundary conditions, which are summarised in section \ref{sec:Three-res}. While for asymptotically flat spacetimes, the metric on the celestial sphere is asymptotically conformal to the round metric, the asymptotically locally flat boundary conditions allow for a general metric on the celestial sphere. 

Previous literature analysed the asymptotics of vacuum Einstein solutions in $d > 4$ assuming asymptotically flat boundary conditions \cite{Tanabe:2009xb,Tanabe:2009va,Tanabe:2010rm,Tanabe2011,Tanabe2012}. In sections \ref{sec:four} and \ref{sec:five} we analyse the asymptotic structure of vacuum Einstein solutions with the weaker, asymptotically locally flat boundary conditions. As in \cite{Hollands2005,Tanabe:2009xb,Tanabe:2009va,Tanabe:2010rm,Tanabe2011,Tanabe2012}, the structure of the expansion depends on whether the spacetime is of odd or even dimension. For definiteness, we focus here on the case of five dimensions, although the generalisation to arbitrary odd and even dimension would be straightforward. 

For five dimensional asymptotically locally flat spacetimes, the asymptotic expansion is polyhomogeneous, i.e. each metric function is expanded as
\be
f(r,u,x^A) = \sum_{i,j} f_{ij} (u,x^A) \frac{\ln^j r}{r^i} 
\ee
where we work in Bondi gauge near future null infinity: $r$ is the radial coordinate, $u$ is the null time coordinate and $x^A$ are the coordinates along the sphere. The coefficients $f_{ij}(u,x^A)$ are smooth functions of their arguments. 

In this paper we do not analyse in detail the asymptotic symmetry group and its implications for soft scattering: we postpone this study for future work. However, it is clear that by relaxing the boundary conditions from asymptotically flat to asymptotically locally flat, the phase space of solutions and the asymptotic symmetry group are necessarily expanded. 

There are striking analogies between the structure of the five dimensional asymptotically locally flat spacetimes we have constructed and that of asymptotically locally anti-de Sitter spacetimes in five dimensions. In both cases the coefficients of the leading logarithm terms are expressed in terms of derivatives of the non-normalizable data (the boundary conditions). In the case of anti-de Sitter, the occurrence of logarithmic terms is associated with conformal anomalies in the dual field theory. 

While much of the earlier relativity literature imposed strictly anti-de Sitter boundary conditions, it is essentially to relax these boundary conditions to asymptotically anti-de Sitter in the context of holography. The generalized boundary condition represents the background metric for the dual field theory. Even if one is only interested in the dual field theory in a flat background, one needs to allow the background metric source to vary to compute correlation functions of the stress energy tensor. 
As we discuss in sections \ref{sec:five} and sections \ref{sec:seven}, it would be interesting to express five-dimensional asymptotically locally anti-de Sitter solutions in Bondi gauge and take the limit of zero cosmological constant, to compare with our results here.  

	\bigskip

	The plan of this paper is as follows. In section \ref{sec:onetwo} we construct cosmic brane solutions in dimensions $d > 4$. In section \ref{sec:Three} we derive the appropriate boundary conditions for vacuum gravity, such that snapping cosmic branes are included in the phase space. In section \ref{sec:four} we write the Einstein equations in Bondi gauge, and then solve these equations subject to asymptotically locally flat boundary conditions. We discuss the structure of the asymptotic expansions in section \ref{sec:five}, give a preliminary analysis of the asymptotic symmetry group in section \ref{sec:six} and we conclude in section \ref{sec:seven}. Appendix \ref{appA} contains a summary of the expansion coefficients of the metric functions while appendix \ref{appB} discusses how the asymptotic analysis can be carried out using iterative differentiation of the equations of motion.

	\section{Defects in dimensions higher than four} \label{sec:onetwo}
	
	In this section we will consider the behaviour of cosmic strings and branes in dimensions higher than four, following the analysis of Vilenkin in four dimensions \cite{Vilenkin:1981zs}; see also discussions in \cite{Vilenkin:2000jqa,Anderson2002}.
	
	Let us consider a $d$-dimensional spacetime, with coordinates $(t,w,\bf{x}$). Suppose a static cosmic string is extended along the $w$ direction, through $\bf{x} =0$; by translation invariance the $\bf{x}$ position can always be chosen to be zero. Let $\mu$ be the tension of the cosmic string. Then the effective stress energy tensor sourcing the cosmic string is \cite{Vilenkin:1981zs}
	\be
	T^{\mu}_{\nu} = \mu \delta^{(d-2)}(x) \;  {\rm diag} (1,1,\bf{0}). \label{cosmict}
	\ee
	Physically, this equation states that the energy density is equal to minus the pressure along the string direction. We will discuss higher dimension defects below. 
	
	\subsection{Linearized gravity}
	
	Now let us consider the backreaction of this stress energy tensor on the spacetime; we assume that $\mu$ is small and thus work within linearized gravity. The $d$-dimensional metric is 
	\be
	g_{\mu \nu} = \eta_{\mu \nu} + h_{\mu \nu}
	\ee
	where $\eta$ is the Minkowski metric and $h$ is the metric perturbation. The Einstein equations can then be expressed as
	\be
	\partial^{\rho} \partial_{\nu} h_{\mu \rho} + \partial^{\rho} \partial_{\mu} h_{\nu \rho} - \Box h_{\mu \nu} - \partial_{\mu} \partial_{\nu} h + (\Box h - \partial^{\rho} \partial^{\sigma} h_{\rho \sigma})\eta_{\mu\nu} = 2 T_{\mu \nu}
	\ee
	where we have set $8 \pi G  = 1$; $\Box$ is the $d$-dimensional d'Alambertian and we define 
	\be
	h = \eta^{\mu \nu} h_{\mu \nu}. 
	\ee
	We impose the usual harmonic gauge
	\be
	\partial^{\nu} h_{\mu \nu} = \frac{1}{2} \partial_{\mu} h.
	\ee
	The remaining gauge invariance is then captured by diffeomorphisms 
	\be
	h_{\mu \nu} \rightarrow h_{\mu \nu} + \partial_{\nu} \xi_{\mu} + \partial_{\mu} \xi_{\nu}
	\ee
	for which 
	\be
	\Box \xi_{\mu} = 0. 
	\ee
	In harmonic gauge the Einstein equations can be expressed as 
	\be
	\Box h_{\mu \nu} = - 2 \tilde{T}_{\mu \nu} \label{ein1}
	\ee
	where $\tilde{T}_{\mu \nu}$ is the trace adjusted stress tensor
	\be
	\tilde{T}_{\mu \nu} = ( T_{\mu \nu} - \frac{1}{(d-2)} T \eta_{\mu \nu} )
	\ee
	with $T = \eta^{\rho \sigma} T_{\rho \sigma}$. 
	
	\subsection{Cosmic strings in $d > 4$}
	
	We now solve the linearized Einstein equation \eqref{ein1} with a trace adjusted stress tensor corresponding to a cosmic string \eqref{cosmict}:
	\be
	\tilde{T}_{\mu \nu} = \frac{\mu}{(d-2)} \delta^{(d-2)}(x) \; {\rm diag} \left ( (4-d), (d-4), - 2 {\bf 1}_{(d-2)} \right ). \label{cosmicttilde}
	\ee 
	Note that the metric backreaction should, by symmetry, be independent of the worldsheet coordinates $(t,w)$ and should be 
	rotationally symmetric in the transverse directions.
	
	In $d=4$ the solution to the linearized equations can be written as  \cite{Vilenkin:1981zs}
	\begin{eqnarray}
	h_{tt} &=& h_{ww} = 0; \\
	h_{xx} &=& h_{yy} = - \tilde{\mu} \ln \left (\frac{r}{r_o} \right ) \nonumber 
	\end{eqnarray} 
	where $r^2 = x^2 + y^2$ and 
	\be
	\tilde{\mu} = \frac{\mu}{2 \pi}.
	\ee
	In this solution $r_o$ can be interpreted as the characteristic radius scale of the string. The linearized solution is valid provided that $| h | \ll 1$, and thus the linearized solution is strictly only applicable within a neighbourhood of the string.  
	
	Thus one can write the four-dimensional (linearized) cosmic string metric as 
	\be
	ds^2 = -dt^2 + dw^2 + (1 - \lambda) (dr^2 + r^2 d \phi^2)
	\ee
	where we use $(r,\phi)$ as polar coordinates in the $(x,y)$ plane and 
	\be
	\lambda = \tilde{\mu} \ln \left ( \frac{r}{r_o} \right ). 
	\ee
	Introducing a new radial coordinate 
	\be
	(1 - \lambda) r^2 = (1 - \tilde{\mu}) \tilde{r}^2
	\ee
	(and working to linear order in $\tilde{\mu}$) one can change the metric into the more familiar form 
	\be
	ds^2 = -dt^2 + dw^2 + d \tilde{r}^2 + (1 - \tilde{\mu}) \tilde{r}^2 d \phi^2
	\ee
	i.e. the metric is locally flat in the directions transverse to the string, but there is a conical deficit proportional to $\tilde{\mu}$. 
	Note that, even though the derivation above was at the level of the linearised equations, this metric manifestly solves the Einstein equations at non-linear order and is moreover locally flat.
	
	\bigskip
	
	Now let us turn to $d > 4$. A qualitative difference in $d \neq 4$ is that the components of \eqref{cosmicttilde} along the string do not vanish. Consider the equation
	\be
	\Box f = \frac{2}{(d-2)} \mu \delta^{(d-2)}(x).
	\ee
	Solutions with rotational symmetry in the directions transverse to the string can be expressed as 
	\be
	f = - \frac{\tilde{\mu}}{r^{(d-4)}}
	\ee
	for $d > 4$ with
	\be
	\tilde{\mu} = \frac{2}{(d-2) \Omega_{(d-3)}} \mu
	\ee
	where $\Omega_{(d-3)}$ is the volume of a unit $(d-3)$-sphere. Then the metric near the cosmic string can be written as 
	\be
	ds^2 = \left ( 1- \frac{(d-4)}{2} \frac{\tilde{\mu}}{r^{(d-4)}} \right ) \left ( -dt^2 + dw^2 \right ) + 
	\left ( 1 - \frac{\tilde{\mu}}{r^{(d-4)}} \right ) \left ( dr^2 + r^2 d \Omega_{(d-3)}^2 \right ).
	\ee
	This solution is not locally Riemann flat close to the cosmic string, although since it is satisfies the Einstein equations (with a string source) 
	it is Ricci flat for $r \neq 0$. The metric is asymptotically locally flat for $r^{d-4} \gg \tilde{\mu}$. However, since 
	an infinite cosmic string necessarily intersects the celestial sphere in two points, and the metric is not locally flat in the immediate neighbourhood of the string, the cosmic string metric is not asymptotically locally flat over the entire celestial sphere. 
	
	\subsection{Cosmic branes}
	
	Let us now consider a $d$-dimensional spacetime, with coordinates $(t,\bf{w},\bf{x}$), where there are $p$ spatial coordinates $\bf{w}$ and correspondingly $(d - p -1)$ transverse coordinates $\bf{x}$. A static cosmic $p$-brane is extended along the $\bf{w}$ directions and located at $\bf{x} =0$. (By translation invariance the $\bf{x}$ position can again always be chosen to be zero.) Let $\mu$ be the tension of the cosmic brane. Then the effective stress energy tensor sourcing the cosmic brane is, generalizing the cosmic string, 
	\be
	T^{\mu}_{\nu} = \mu \delta^{(d-p-1)}(x) \;  {\rm diag} (1,{1}_{(p)},{0}_{(d-p-1)}). \label{cosmicbt}
	\ee
	Physically, this equation states that the energy density is equal to minus the pressures along the brane. Note that in four dimensions a cosmic membrane would usually be referred to as a domain wall, as there is only one transverse direction, and such solutions were discussed together with cosmic strings in \cite{Vilenkin:1981zs}.
	
	The corresponding trace adjusted stress tensor is then 
	\be
	\tilde{T}_{\mu \nu} = \frac{\mu}{(d-2)} \delta^{(d-p-1)}(x) \;  {\rm diag} (- (d - p - 3),{ (d-p-3)}_{(p)},- { (p+1)}_{(d-p-1)}).
	\ee
	In the case that $d = (p + 3)$ this implies that 
	\be
	h_{tt} = h_{ww} = 0,
	\ee
	i.e. the metric perturbations longitudinal to the brane are zero. The transverse space to the brane then has dimension two and the corresponding form for the metric near the cosmic brane is 
	\be
	ds^2 = - dt^2 + d w \cdot dw_{(d-3)} + (1 - \lambda) (dr^2 + r^2 d \phi^2) \label{c-brane}
	\ee
	where now
	\be
	\lambda =  \tilde{\mu} \ln \left (\frac{r}{r_o} \right )
	\ee
	with 
	\be
	\tilde{\mu} = \frac{(p+1)}{(d-2)} \frac{\mu}{2 \pi}.
	\ee
	Following the same logic as above, the metric \eqref{c-brane} can be written in a form which is manifestly locally flat, namely
	\be
	ds^2 = - dt^2 + d w \cdot dw_{(d-3)} + d\tilde{r}^2 + (1 - \tilde{\mu}) \tilde{r}^2 d \phi^2, \label{c2-brane}
	\ee
	with the transverse space to the brane having a conical singularity at $\tilde{r} = 0$. 
	
	In the case that $d \neq (p+3)$ the metric perturbations longitudinal to the brane are non-zero, and the solution near the brane is Ricci flat but not locally Riemann flat, as in the case of cosmic strings in $d \neq 4$ discussed above. Thus we see that branes of codimension two play a distinguished role when we are interested in asymptotically locally flat geometries. 
	
	\bigskip
	
We should note that there has been considerable discussion in the relativity literature about distributional sources. The analysis of \cite{Geroch:1987qn} highlighted subtleties in dealing with distributional source of codimension greater than one: the metric is inherently distributional and the curvature is constructed from products of metric derivatives. This implies that different regularisations of cosmic strings can lead to thin, static strings with different mass per unit lengths. Later work by Garfinkle \cite{Garfinkle:1999xv} defined a notion of semi-regular metrics, in which the static cosmic string has a distributional stress energy; however, it is also argued in this work that such stress energy may not actually describe the physical energy content. The work of \cite{Traschen:2008cq} explored distributional brane sources, showing that one can make sense of stress confined to codimension two surfaces in certain situations. There is also an ongoing programme of work using generalized functions to understand distributional curvature, beginning with \cite{Clarke:1996pp,Wilson:1997xk}. 

Following the earlier work of \cite{Strominger2016}, the metric \eqref{c2-brane} will be the starting point for our analysis, and motivation for considering more general boundary conditions than asymptotically flat in $d > 4$.  The detailed description of the distributional curvature will not be central to our analysis. Ultimately the physical interpretation of such branes may well go beyond general relativity into string theory, in which branes are valid physical objects with well understood stress energy (and where the limits of the validity of general relativity solutions are also understood).

	\subsection{Cosmic branes: general position and orientation}
	
	In the previous section we gave solutions that are longitudinal to the $\bf{w}$ directions and located at the origin in the transverse directions. 
	It is clearly straightforward to generalize such solutions to arbitrary position and orientation. Let us first write the solution \eqref{c2-brane} in terms of Cartesian coordinates in the transverse directions i.e. 
	\begin{eqnarray}
	d\tilde{r}^2 + (1 - \tilde{\mu}) \tilde{r}^2 d \phi^2 &=& \frac{1}{(x^2 + y^2)} \left ( (x^2 + K^2 y^2) dx^2 + 2 x y (1 - K^2) dx dy \right . \label{cart} \\
	&& \qquad \qquad \qquad \left .  + (y^2 + K^2 x^2) dy^2 \right ), \nonumber \\
	&=& dx^2 + dy^2 - \frac{\tilde{\mu}}{(x^2 + y^2)} \left (y dx - x dy \right )^2 \nonumber
	\end{eqnarray}
	where we use the notation $K^2 = (1 - \tilde{\mu})$. Clearly when $K=1$ this metric reduces to the standard Euclidean metric in Cartesian coordinates. Note that one can also write \eqref{c2-brane} as
	\be
	ds^2 =  dz d \bar{z} + \frac{\tilde{\mu}}{4 z \bar{z}} \left ( \bar{z} dz - z d \bar{z} \right )^2 
	\ee
	in terms of a complex coordinate $z = (x + i y)$.
	
	It is then straightforward to displace the brane from the origin to $(x_o,y_o)$ by shifting
	\be
	x \rightarrow (x - x_o) \qquad y \rightarrow (y - y_o). 
	\ee
	Clearly for $K = 1$ this would leave the metric invariant but for general $K$ we obtain
	\be
	dx^2 + dy^2 - \frac{\tilde{\mu}}{((x-x_o)^2 + (y-y_o)^2)} \left ( (y-y_o) dx - (x-x_o) dy \right )^2. 
	\ee
	Moving back to polar coordinates, the metric takes the simple form
	\be
	d \tilde{r}^2 + K^2 (\tilde{r} - \tilde{r}_o)^2 d \phi^2 \label{displace}
	\ee
	where $\tilde{r}_{o}^2 = (x_{o}^2 + y_o^2)$.
	
	\bigskip
	
	Using the Cartesian form of the metric \eqref{cart} it is also straightforward to rotate the orientation of the brane. For example, if we rotate in the $(wx)$ plane by an angle $\alpha$ via
	\be
	w \rightarrow \cos \alpha w - \sin \alpha x \qquad
	x \rightarrow  \cos \alpha x + \sin \alpha w
	\ee
	we obtain
	\be
	dw^2 + dx^2 + dy^2 - \frac{\tilde{\mu}}{( (\cos \alpha x + \sin \alpha w)^2 + y^2)} \left ( y (\cos \alpha d x + \sin \alpha  d w) - (
	\cos \alpha x + \sin \alpha w) dy \right )^2
	\ee
	Note that the residual rotational symmetry transverse to the brane is not manifest in this coordinate system. 
	
	\subsection{Asymptotics}
	
	Let us now return to  \eqref{c2-brane}. To analyse the asymptotics we should rewrite it as 
	\be
	ds^2 = - dt^2 + dw^2 + w^2 d \Omega_{p-1}^2 + d \tilde{r}^2 + K^2 \tilde{r}^2 d \phi^2
	\ee
	and then let 
	\be
	w = R \cos \Theta \qquad
	\tilde{r} = R \sin \Theta
	\ee
	to obtain
	\be
	ds^2 = -dt^2 + dR^2 + R^2 \left ( d \Theta^2  + \cos^2 \Theta d \Omega_{p-1}^2 + K^2 \sin^2 \Theta d \phi^2 \right ).
	\ee
	Note that both the longitudinal $SO(p)$ rotational symmetry and the transverse $SO(2)$ rotational symmetry are manifest. 
	
	\bigskip
	
	A brane which is located at $\tilde{r} = \tilde{r}_o$ \eqref{displace} can be expressed as
	\begin{eqnarray}
	ds^2 &=& -dt^2 + dR^2 + R^2 \left ( d \Theta^2  + \cos^2 \Theta d \Omega_{p-1}^2 + \sin^2 \Theta d \phi^2 \right ) \\
	&& - \tilde{\mu} \frac{\left ( - R^2 \sin^2 \Theta  d \phi -  \tilde{r}_o \sin (\phi - \phi_o) d (R \sin \Theta) + \tilde{r}_o R \sin \Theta \cos (\phi - \phi_o) d \phi \right )^2}
	{(R^2  \sin^2 \Theta  + \tilde{r}_o^2 - 2 \tilde{r}_o R \sin \Theta \cos (\phi - \phi_o) )} \nonumber
	\end{eqnarray}
	or equivalently as 
	\begin{eqnarray}
	ds^2 &=& -dt^2 + dR^2 + R^2 \left ( d \Theta^2  + \cos^2 \Theta d \Omega_{p-1}^2 + \sin^2 \Theta d \phi^2 \right ) \label{dis1} \\
	&& - \tilde{\mu} \frac{\left ( - R \sin^2 \Theta  d \phi -  \sin \Theta_o \sin (\phi - \phi_o) d (R \sin \Theta) + R \sin \Theta_o \sin \Theta \cos (\phi - \phi_o) d \phi \right )^2}
	{(\sin^2 \Theta  + \sin^2 \Theta_o - 2 \sin \Theta_o \sin \Theta \cos (\phi - \phi_o) )} \nonumber
	\end{eqnarray}
	where we define
	\be
	\sin \Theta_o = \frac{\tilde{r}_o}{R}. 
	\ee
	For $\Theta_o$ to remain finite as $R \rightarrow \infty$ we will clearly need to take $\tilde{r}_o$ to infinity with the ratio of $\tilde{r}_o/R$ fixed. 
	
	Note that the metric \eqref{dis1} has a hidden $U(1)$ symmetry, corresponding to rotations around $\tilde{r} = \tilde{r}_o$. 
	The metric on a surface of constant $R$ and $t$ is
	\begin{eqnarray}
	ds^2 &=& 
	R^2 \left ( d \Theta^2  + \cos^2 \Theta d \Omega_{p-1}^2 + \sin^2 \Theta d \phi^2 \right ) \label{dis2}\nn \\
	&& - \tilde{\mu} R^2 \frac{\left ( \sin^2 \Theta   +  \sin \Theta_o \sin (\phi - \phi_o  - \Theta) \right )^2 d \phi^2}
	{(\sin^2 \Theta  + \sin^2 \Theta_o - 2 \sin \Theta_o \sin \Theta \cos (\phi - \phi_o) )}\,\;. 
	\end{eqnarray}
	A surface of constant $R$ and $t$ clearly does not have such a $U(1)$ symmetry; only the $SO(p)$ symmetry along the longitudinal directions of the brane survives. This is illustrated in the case of $p=1$ in Figure~\ref{fig:1a}: the string clearly has an axial $SO(2)$ symmetry but the intersection with the celestial two sphere does not preserve this $SO(2)$ symmetry. 
	
	\begin{figure}[tbp]
		\centering
		\includegraphics[width=0.4\linewidth]{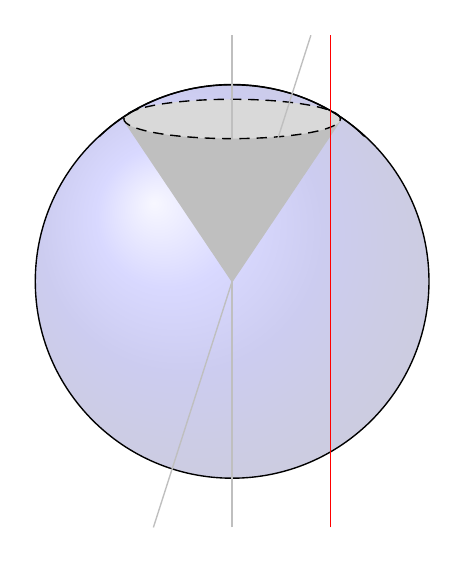}
		\caption{Three cosmic strings: a string passing though the north pole of the sphere; a string rotated with respect to this axis and a third string (red) translated with respect to the first one.}
		\label{fig:1a}
	\end{figure}
	
	The asymptotics of a rotated cosmic brane can also be obtained using the radial coordinate $R$. In this case there is an axial 
	$SO(2)$ symmetry which is respected by the intersection with the celestial sphere; this is however not manifest in the coordinates 
	$(\Theta, \phi)$. A rotated string is shown in Figure~\ref{fig:1a}. 
	
	\bigskip
	
	Note that much of the previous literature on 5d cosmic branes has concentrated on spacetimes with cylindrical symmetry i.e. one writes the metric for flat space as 
	\be
	ds^2 =  -du^2 - 2 du d\rho + \rho^2 (d \theta^2 + \sin^2 \theta d \phi^2) + dz^2 \label{cosmic3a}
	\ee
	i.e. as a direct product of four-dimensional Minkowski spacetime with a line. This form of the metric is particularly convenient when one compactifies the $z$ direction around a circle i.e. one is interested in a Kaluza-Klein spacetime or a brane world Randall-Sundrum setting. However, \eqref{cosmic3a} is not expressed in a form that is natural for analysing the asymptotic structure if $z$ is not compact; analysis of the structure close to null infinity requires the introduction of a radial coordinate 
	$r^2 = \rho^2 + z^2$, to characterise the celestial sphere.

	\section{Cosmic branes and asymptotically locally flat spacetimes} \label{sec:Three}
	
	In this section we consider the asymptotic structure of cosmic $(d-3)$-branes and show how such spacetimes can be expressed in 
	Bondi gauge. This analysis demonstrates the boundary conditions that should be imposed 
	on the metric functions in Bondi gauge so that cosmic $(d-3)$-branes are contained within the set of solutions of the vacuum Einstein 
	equations. 
	
	For $d > 4$, the boundary conditions required are weaker than those imposed in earlier literature: the inclusion of cosmic $(d-3)$-branes defines boundary conditions for asymptotically locally flat spacetimes. For concreteness we focus mostly on the case of $d=5$ but the generalization of these boundary condtions to arbitrary $d > 4$ is straightforward and is summarised at the end of this section.

	\subsection{Cosmic strings in four dimensions} \label{sec:two}
	
	Let us begin by reviewing relevant features of cosmic strings in four dimensions. We can write the metric for four-dimensional Minkowski spacetime in coordinates adapted to future null infinity ${\mathscr{I}}^+$ as
	\be
	ds^2 = - du^2 - 2 du dr + r^2 (d \theta^2 + \sin^2 \theta d \phi^2). \label{metric1}
	\ee
	Now let us suppose we want to allow for dynamical processes in which there is a transition from this vacuum spacetime to a cosmic string spacetime. A static cosmic string metric can be written as 
	\be
	ds^2 = - dU^2 -  2 d U dR + R^2 (d \Theta^2 + K^2 \sin^2 \Theta d \Phi^2) \label{metric2}
	\ee                                                                                                                                                                                                                                                                                                                                                                                                                                                                                                                                                                                                                                                                                                                                                                                   
	where $K^2 = 1 - 2 \delta$ characterizes the deficit angle. Note that the cosmic string intersects the celestial sphere at the north and south pole i.e. $\Theta = 0, \pi$. 
	
	From the perspective of ${\mathscr{I}}^+$, a process in which a cosmic string is destroyed manifests as a transition from a metric on null  hypersurfaces that has a deficit, to a metric that is a round sphere. Following the discussions in \cite{Bicak1989,Podolsky2000}, we can match \eqref{metric1} and \eqref{metric2}
	on a null hypersurface at large $r$:
	\be
	R^2 (d \Theta^2 + K^2 \sin^2 \Theta d \Phi^2) = r^2 (d \theta^2 + \sin^2 \theta d \phi^2) + \cdots \label{matching}
	\ee
	where the ellipses denote terms that are subleading in $r$. By symmetry one can identify $\Phi = \phi$. The other coordinate transformations admit asymptotic expansions 
	\begin{eqnarray}
	U &=& U_{(0)} (u, \theta) + \frac{U_{(-1)}(u, \theta)}{r} + \cdots \\
	R &=& r R_{(1)} (\theta) + R_{(0)} (u,\theta)  + \cdots \nonumber \\
	\Theta &=& \Theta_{(0)} (\theta) + \frac{\Theta_{(-1)} (\theta)}{r} + \cdots \nonumber
	\end{eqnarray}
	Here $R_{(1)}(\theta)$ and $\Theta_{(0)}(\theta)$ are necessarily independent of $u$ to preserve the leading radial dependence of the $u$ components of the metric. The transformation $U_{(0)}(u,\theta)$ is not determined by the leading order matching of \eqref{matching}, which imposes
	\begin{equation}
	R_{(1)}^2 (\partial_{\theta} \Theta_{(0)})^2 = 1; \qquad
	K^2 R_{(1)}^2 \sin^2 \Theta_{(0)} = \sin^2 \theta.
	\end{equation}
	These equations can be integrated to give 
	\begin{equation}
	R_{(1)} = \frac{\sin \theta}{K \sin \Theta_{(0)}}; \qquad
	\int \mathrm{cosec} \Theta_{(0)} \; d \Theta_{(0)} = K \int \mathrm{cosec}  \theta \; d \theta. \label{4dsol}
	\end{equation}
	Note that the transformation is not analytic as $\theta \rightarrow 0$; further discussions can be found in \cite{Bicak1989,Podolsky2000}. 
	
	This analysis shows that in four dimensions one can match a cosmic string metric as $r \rightarrow \infty$ with the Minkowski metric, at leading order. After changing coordinates to match at leading order, the subleading terms in the cosmic string metric are non-zero, see \cite{Podolsky2000,Strominger2017}; in \cite{Strominger2017} these subleading terms were interpreted as superrotations \cite{Barnich2010}. 
	
	\subsection{Five dimensional cosmic brane metrics}
	
	For the metric in the vicinity of the brane to be locally flat, the brane must be a $(d-3)$-brane, i.e. a membrane in five dimensions.
	In five dimensions we can parameterise locally flat metrics with deficits in several ways and in this section we discuss convenient parameterisations. 
	
	Let us first consider 
	\be
	ds^2 = - dU^2 - 2 dU dR + R^2 (d \Theta^2 + \cos^2 \Theta d \Psi^2 + K^2 \sin^2 \Theta d \Phi^2). \label{cosmic5d1}
	\ee
	For $K^2=1$ hypersurfaces of constant $U$ are round three spheres, with a $U(1)^2$ subgroup of the $SO(4)$ isometry group made manifest. If we introduce a deficit $K^2 = 1 - 2 \delta$, the deficit is associated with $\Theta = 0$, but extends around the entire $\Psi$ circle i.e. there is a cosmic membrane intersecting the celestial three-sphere in a circle. To see this, it is convenient to exploit the embedding of the three sphere into $R^4$ i.e. 
	\be
	x = R \cos \Theta \cos \Psi; \; y = R \cos \Theta \sin \Psi; \;
	z = R \sin \Theta \cos \Phi; \;
	w = R \sin \Theta \sin \Phi.
	\ee
	Thus $\Theta = 0$ corresponds to the circle $x^2 + y^2 = R^2$ with $z = w = 0$. 
	
	There is an obvious generalisation of \eqref{cosmic5d1}:
	\be
	ds^2 = - dU^2 - 2 dU dR + R^2 (d \Theta^2 + K_1^2 \cos^2 \Theta d \Psi^2 + K^2_2 \sin^2 \Theta d \Phi^2) \label{cosmic5d1a}
	\ee
	in which for $K_1^2 \neq 1$ and $K_2^2 \neq 1$ there is a cosmic membrane intersecting the sphere in the circle $x^2 + y^2 = R^2$ with $z = w = 0$ and a second membrane intersecting $z^2 + w^2 = R^2$ with $x = y  = 0$. This specific configuration of membranes preserves the $U(1)^2$ symmetry associated with rotations in the $(x,y)$ and $(w,z)$ planes. 
	
	\bigskip
	
	We could alternatively study
	\be
	ds^2 = - dU^2 - 2 dU dR + R^2 (d \Theta^2 + K_1^2 \sin^2 \Theta d X^2 + K_2^2 \sin^2 \Theta \sin^2 (K_1 X ) d \Phi^2) \label{cosmic5d2}
	\ee 
	where for $K_1^2 = K_2^2 = 1$ hypersurfaces of constant $U$ are round three spheres, in which an $SO(3)$ subgroup of $SO(4)$ is made manifest. The metric is manifestly locally flat for $K_1 \neq 1$ and 
	$K_2 \neq 1$: this follows from the coordinate redefinitions $\chi = K_1 X$ and $\phi = K_2 \Phi$, which bring the metric into the form of a flat metric. These coordinate redefinitions are locally trivial; deficits are introduced by imposing the standard ranges on the redefined coordinates i.e.
	$0 \le X < \pi$ and $0 \le \Phi < 2 \pi$.  
	
	When a deficit is introduced by setting $K_2^2 \neq 1$ (with $K_1^2 = 1$), the deficit is associated with $X = 0, \pi$. The interpretation is again most easily seen by embedding the (round) three sphere into ${\rm R}^4$ as 
	\be
	x = R \sin \Theta \sin X \sin \Phi; \qquad
	y = R \sin \Theta \sin X \cos \Phi; \qquad
	z = R \sin \Theta \cos X; \qquad
	w = R \cos \Theta 
	\ee
	i.e. the deficit is associated with $z^2 + w^2 = R^2$, $x =  y = 0$, a great circle of the sphere. This metric thus describes the same physics as the metric shown in \eqref{cosmic5d1} but the parameterisation of \eqref{cosmic5d2} is less convenient, as it does not make manifest the second $SO(2)$ symmetry preserved by the cosmic membrane.  
	
	When $K_1^2 \neq 1$ (with $K_2^2 = 1$) the deficit is associated with geodesic incompleteness of the two spheres parameterized by $(X,\phi)$. For constant $U$, $R$ and $\Theta$ the induced two-dimensional metric is
	\be
	ds^2 = R^2 \sin^2 \Theta K_1^2 (dX^2 +  \frac{1}{K_1^2} \sin^2 (K_1 X) d \Phi^2) \label{2dd}
	\ee
	which describes part of a two sphere of radius $K_1 R \sin \theta$; more precisely, since $0 \le X > \pi$, there is a boundary to \eqref{2dd} at $X = \pi$:
	\be
	ds^2 = R^2 \sin^2 \Theta \sin^2 (K_1 \pi) d \Phi^2
	\ee
	i.e. a circle. We will not consider this case further as it does not seem to have a natural physical interpretation. 
\begin{figure}[tbp]\centering\label{fig:1}
\includegraphics*[width=0.4\linewidth]{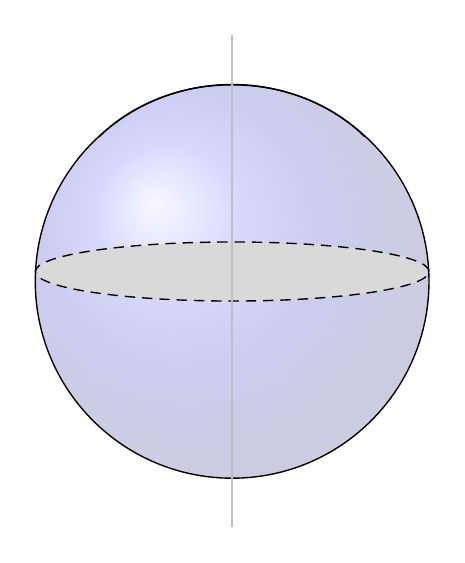}
	\caption{Cosmic string and membrane intersecting the celestial sphere.}
\end{figure}
	

	Let us now relate these discussions to the cosmic brane solutions of the previous section. The metric \eqref{cosmic5d2} can be written in terms of a time coordinate
	\be
	t = U + R
	\ee
	as
	\be
	ds^2 = -dt^2 + dR^2 + R^2 ( d \Theta^2 + K_1^2 \cos^2 \Theta d \Psi^2 + K_2^2 \sin^2 \Theta d \Phi^2).
	\ee
	Now introduce coordinates
	\be
	\tilde{r} = R \cos \Theta; \qquad
	w = R \sin \Theta
	\ee
	in terms of which the metric can be expressed as 
	\be
	ds^2 = -dt^2 + ( d \tilde{r}^2 + K_1^2 \tilde{r}^2 d \Phi^2) + (dw^2 + K_2^2 w^2 d \Psi^2)
	\ee
	Consider first the case of $K_2^2  =1$. Comparing with \eqref{c2-brane}, the cosmic membrane is 
	located at $\tilde{r}=0$, i.e. $\Theta = \Pi/2$, and lies in the $(w,\Psi)$ plane. This defect is visualised in Figure~\ref{fig:1}, as the 
	plane intersecting the celestial sphere in a circle. For $K_2^2 \neq 1$, there is in addition a membrane located at $w = 0$, lying in the $(\tilde{r}, \Phi)$ plane. This second membrane intersects the first at $\tilde{r} = w = 0$. Note that this intersection does not take place close to the celestial sphere, so any non-linear effects at the intersection are not relevant for asymptotic analysis.  
	
	\bigskip
	
	Before we move to the general asymptotic analysis, let us consider an infinite cosmic string, which as shown in Figure~\ref{fig:1} necessarily intersects the celestial sphere at two points. We are interested in metrics which are asymptotically locally flat at infinity. However, from the discussions of the previous section, a cosmic string metric is only locally flat (as opposed to Ricci flat) near the string in four dimensions. We therefore cannot match a cosmic string metric with a flat metric on a null hypersurface, except in four dimensions; equivalently, we cannot apply coordinate transformations to a flat metric and obtain a cosmic string metric in dimensions other than four.

	\subsection{Asymptotically locally flat metrics in five dimensions}
	
	Now let us extend the discussion of section \ref{sec:two} to five dimensions. We allow for dynamical processes in which cosmic branes are created and destroyed and, as before, we consider the matching of a cosmic brane metric to a Ricci flat metric without cosmic brane on a null hypersurface. 
	
	For computational simplicity we consider cosmic brane metrics that preserve $U(1)^2$ symmetry in the angular directions and have reflection/inversion symmetry in these directions. Such metrics can be matched to asymptotically locally flat spacetimes with corresponding symmetry which can be described using a Bondi gauge parametrisation:
	\begin{eqnarray}
	ds^2 &=& - ({\cal U} e^{2 \beta} - r^2 W^2 e^{C_1}) du^2 - 2 e^{2 \beta} du dr - 2 r^2 W e^{C_1} du d \theta \label{bondi1} \nn\\
	&& + r^2 (e^{C_1} d \theta^2 + e^{-(C_1+ C_2)} \cos^2 \theta d \psi^2 + e^{C_2} \sin^2 \theta d \phi^2 ).
	\end{eqnarray}
	Here the defining metric functions $({\cal U},W,\beta,C_1,C_2)$ depend only on $(u,r,\theta)$ due to the symmetry. We have also imposed the standard Bondi gauge conditions i.e. 
	\be
	g_{rr} = g_{rA} = 0 
	\ee
	and the determinant of the angular part of the metric is $r^6$; these conditions mirror the original four-dimensional conditions 
	\cite{Bondi1962, Sachs1962}.  
	
	The standard definition of asymptotically flat spacetimes in five dimensions (see \cite{Hollands2003,Hollands2004,Hollands2005,Tanabe:2009xb,Tanabe:2009va,Tanabe:2010rm,Tanabe2012})
	imposes the following boundary conditions on the defining functions for solutions of the vacuum Einstein equations:
	\begin{eqnarray}
	{\cal U} (u,r,\theta)  &=& 1 + \frac{ {\cal U}_{(3/2)}(r,\theta)}{r^{\frac{3}{2}}} + \cdots  \label{afbc} \\
	W (u,r,\theta) &=& \frac{W_{(3/2)}(r,\theta)}{r^{\frac{3}{2}}} + \cdots \nonumber \\
	{ \beta} (u,r,\theta)  &=& \frac{ {\beta}_{(3)}(r,\theta)}{r^3} + \cdots \nonumber \\
	C_{i} (u,r,\theta) &=& \frac{C_{i (3/2)}(r,\theta)}{r^{\frac{3}{2}}} + \cdots \nonumber
	\end{eqnarray}
	where $i =1,2$ and the ellipses denote terms that are subleading as $r \rightarrow \infty$. As we review below, gravitational waves are associated with the $C_{i (3/2)}$ contributions, which in turn induce subleading terms in the other metric functions. Additional integration functions arise at order $1/r^2$ and are associated with mass and angular momentum; we will discuss these later, when we derive the asymptotic expansions to all orders.  
	
	\bigskip
	
	We now consider the matching  between \eqref{cosmic5d1a} and \eqref{bondi1} on a null hypersurface as $r \rightarrow \infty$ and show that such a matching requires weaker boundary conditions than asympotically flat boundary conditions \eqref{afbc}. 
	By symmetry, we can identify $\Psi = \psi$ and $\Phi = \phi$. Following the four-dimensional discussion, we parameterise the coordinate transformations as
	\begin{eqnarray}
	U &=& U_{(0)} (u,\theta) + \frac{U_{(- 1)}(u,\theta)}{r} + \cdots \label{trans1} \\
	R &=& r R_{(1)}(\theta) + R_{(0)}(u,\theta) + \cdots \nonumber \\
	\Theta &=& \Theta_{(0)} (\theta) + \frac{\Theta_{(-1)}(\theta)}{r} + \cdots \nonumber
	\end{eqnarray}
	Matching on a null hypersurface then imposes three relations at leading order:
	\begin{eqnarray}
	( \theta \theta ): \qquad R_{(1)}^2 (\partial_{\theta} \Theta_{(0)})^2 &=& e^{c_{1}(\theta)} \label{leading1} \\
	( \psi \psi): \qquad R_{(1)}^2 \cos^2 \Theta_{(0)} &=& e^{-(c_1(\theta)+ c_2 (\theta))} \cos^2 \theta \nonumber \\
	(\phi \phi): \qquad R_{(1)}^2 \sin^2 \Theta_{(0)} &=& \frac{1}{K^2} e^{c_2 (\theta)} \sin^2 \theta \nonumber
	\end{eqnarray}
	where we indicate the components of the induced metric being matched and we expand the defining metric functions $(C_1,C_2)$ as
	\be
	C_{i} (u,r,\theta) = c_{i}(\theta) + \frac{C_{i (- \lambda)}(u,\theta)}{r^{\lambda}} + \cdots \label{weaker}
	\ee
	where the exponent $\lambda > 0$ will follow from imposing Ricci flatness.  In the case that $c_i = 0$, then $\lambda = \frac{3}{2}$ as in \eqref{afbc} but this is no longer true when $c_{i} \neq 0$, as we will show below. 
	
	Before we consider the solution of \eqref{leading1}, let us discuss the structure of the coordinate transformations in \eqref{trans1}. As in four dimensions, the leading terms in $R(u,r,\theta)$ and $\Theta(u,r,\theta)$ are forced to be independent of $u$, as $u$ dependence would induce metric components along the $u$ direction that scale as a positive power of $r$, thus breaking the notion of asymptotic local flatness.  
	
	The leading order contributions to the other metric components are:
	\begin{eqnarray}
	(r \theta) &:& {\cal O}(r^0) \qquad g_{r \theta} = R_{(1)} \Theta_{(-1)} \partial_{\theta} \Theta_{(0)} + \partial_{\theta} U_{(0)} = 0 \label{leading2} \\
	(u r ) &:& {\cal O}(r^0) \qquad g_{ur} = R_{(1)} \partial_u U_{(0)} \nonumber \\
	(u u ) &:& {\cal O}(r^0) \qquad g_{uu} = - (\partial_u U_{(0)} )^2 + R_{(1)}^2 (\partial_u \Theta_{(-1)})^2 \nonumber \\
	(u \theta) &:& {\cal O}(r) \qquad g_{u \theta} = \partial_{\theta} R_{(1)} \partial_u U_{(0)} - R_{(1)}^2 \partial_{\theta} \Theta_{(0)} \partial_u \Theta_{(-1)} \nonumber \\
	(r r ) &:& {\cal O}(r^2) \qquad g_{rr} =  2 R_{(1)} U_{(-1)} + R_{(1)}^2 (\Theta_{(-1)})^2 = 0 \nonumber
	\end{eqnarray}
	These relations put no further conditions on $(\Theta_{(0)},R_{(1)})$, which are determined by \eqref{leading1}, but instead determine $(U_{(0)},\Theta_{(-1)},\cdots)$ in terms of these functions. 
	
	\bigskip
	
	Now let us consider the solution of \eqref{leading1}. If we impose strict asymptotic flatness as in \eqref{afbc}, then we need to set $c_1 (\theta) = c_2 (\theta) = 0$. However, in this case the three conditions of \eqref{leading1} are clearly incompatible: the first and third relations are identical to those in four dimensions and are solved as in \eqref{4dsol} but this solution is not consistent with the second relation in \eqref{leading1}. 
	
	One can conceptualise why a three sphere with a ring of conical deficits cannot be mapped to a round three sphere as follows. Hypersurfaces of constant $\Theta$ are topologically tori, with the $\psi$ and $\phi$ circles parameterising the independent non-contractable cycles of these tori. There is a geometric interpretation of solving the first and third relations in \eqref{leading1} (with $c_1 (\theta) = c_2 (\theta ) = 0$): one uses an angle dependent rescaling of the radius to remove the deficit in the $\phi$ circle as $\theta \rightarrow 0$. However, this same angle dependent rescaling of the radius is then incompatible with maintaining the periodicity of the $\psi$ circle. 
	
	We thus conclude that we cannot solve \eqref{leading1} without allowing for non-zero $c_i(\theta)$. However,  if the functions $c_i(\theta)$ are non-zero, 
	the system of equations now seems to be under-constrained: there are only three equations for four functions $(\Theta_{(0)}(\theta),R_{(1)}(\theta),c_1(\theta),c_2(\theta))$. Combining the three equations one can obtain the following relation
	\be
	\frac{d \Theta_{(0)}}{\sqrt{\sin 2 \Theta_{(0)}}} = K^{\frac{1}{2}} \frac{ e^{\frac{3 c_1 (\theta)}{4}} d \theta}{\sqrt{\sin 2 \theta }}.
	\ee
	From this relation we can see that one of the four functions follows from the freedom to redefine the angular coordinate; imposing $\Theta_{(0)} = \theta$ for $K=1$ fixes $c_1(\theta) = 0$.  Thus $\Theta_{(0)}(\theta)$ is given by
	\be
	\int \frac{d \Theta_{(0)}}{\sqrt{\sin 2 \Theta_{(0)}}} = K^{\frac{1}{2}} \int \frac{ d \theta}{\sqrt{\sin 2 \theta }}, \label{theta1}
	\ee
	and the other functions are determined by the relations
	\begin{eqnarray}
	R_{(1)} (\theta) &=& \frac{1}{K^{\frac{1}{2}}} \sqrt{ \frac{\sin 2 \theta}{\sin 2 \Theta_{(0)}}} \label{R1c2} \\
	e^{c_2(\theta)} &=& K \frac{ \tan \Theta_{(0)}}{\tan \theta}. \nonumber
	\end{eqnarray}
	
	\begin{figure}[tbp]
		\centering
		\includegraphics[width=0.5\linewidth]{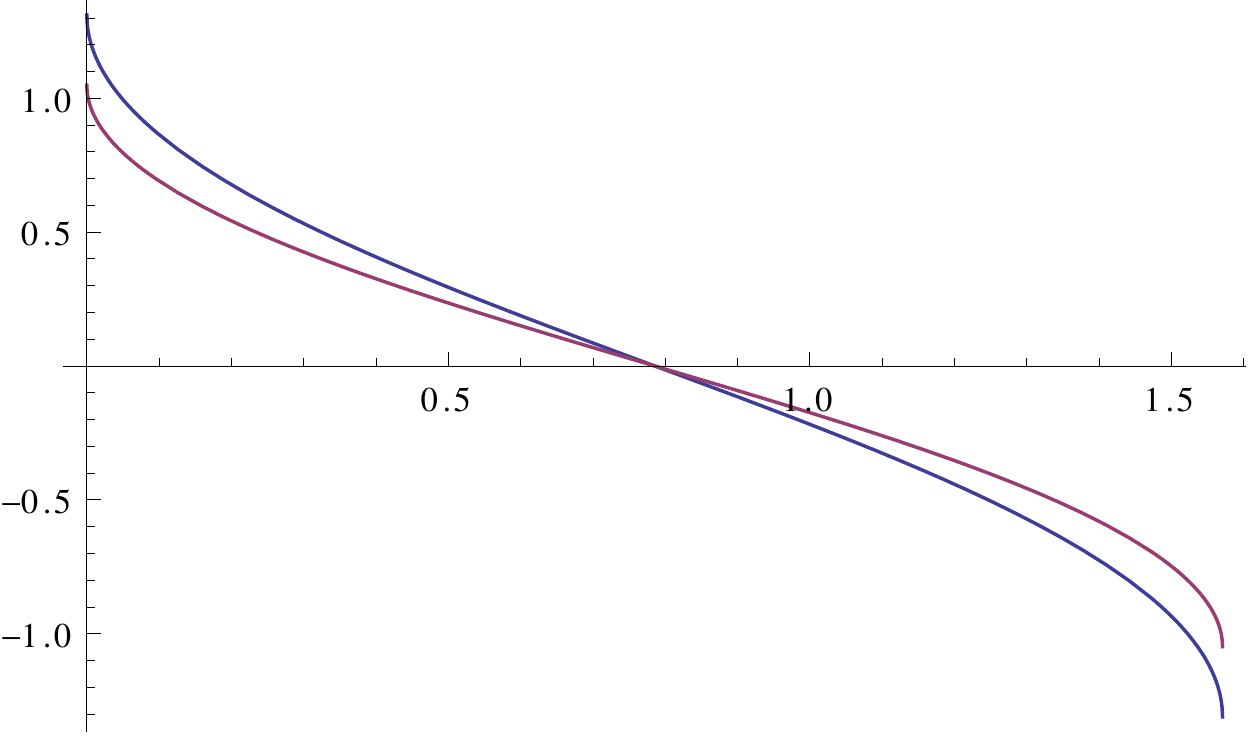}
		\caption{The blue line shows $F \left ( \frac{1}{2} (\frac{\pi}{2} - 2 x) | 2 \right ) $, plotted over the range $(0,\pi/2)$. The red line shows $K^{\frac{1}{2}} F \left ( \frac{1}{2} (\frac{\pi}{2} - 2 x) | 2 \right )$, plotted over the same range, with $K^{\frac{1}{2}} = 0 .8$.}
		\label{fig:1a}
	\end{figure}
	
\begin{figure}[tbp]
		\centering
		\includegraphics[width=0.5\linewidth]{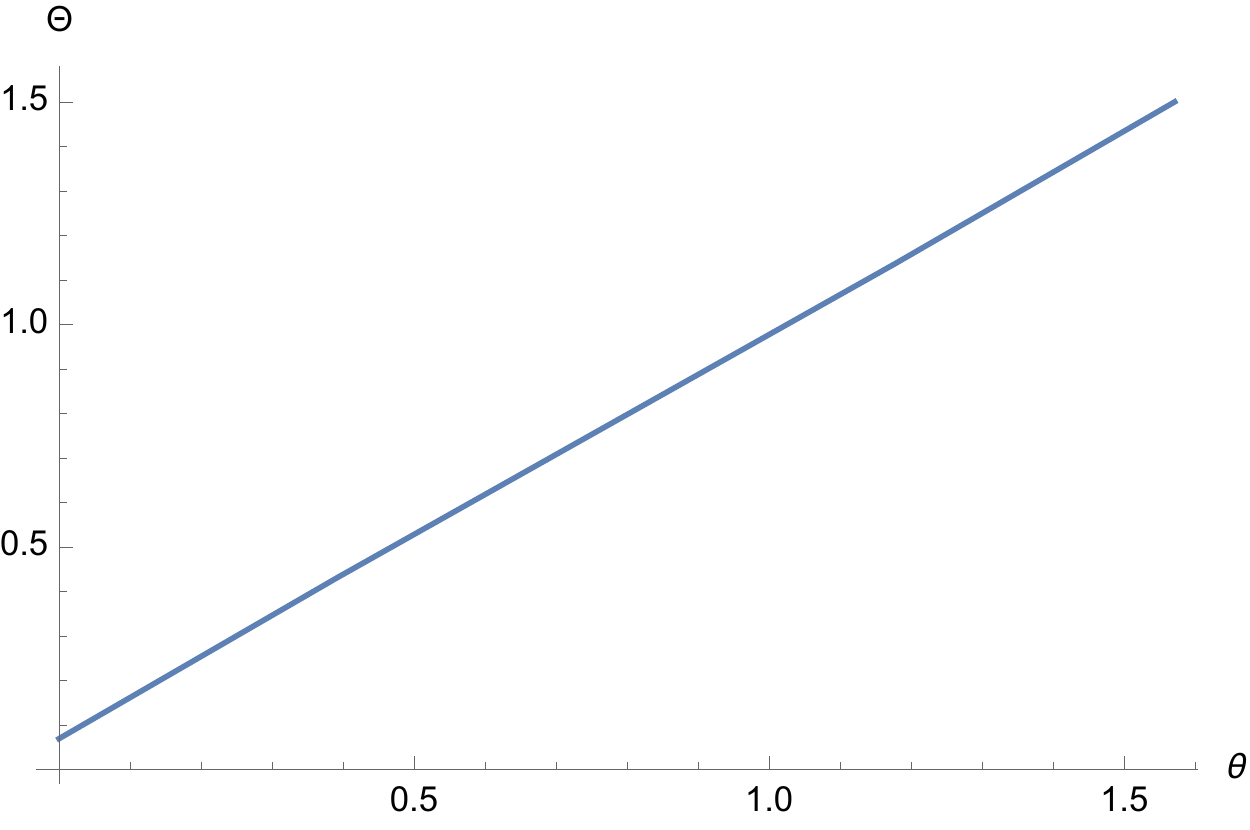}
		\caption{$\Theta_{(0)}(\theta)$ for $K^{\frac{1}{2}} = 0.8$.}
		\label{fig:2}
	\end{figure}

	Integrating \eqref{theta1} we obtain
	\be
	F \left ( \frac{1}{2} (\frac{\pi}{2} - 2 \Theta_{(0)}) | 2 \right ) = K^{\frac{1}{2}} F \left ( \frac{1}{2} (\frac{\pi}{2} - 2 \theta) | 2 \right ), \label{solu1}
	\ee
	where $F(y | m)$ is the elliptic integral of the first kind. This elliptic integral is plotted in Figure~\ref{fig:1a}. For $K^2$ just less than one, we can read off from Figure~\ref{fig:1a} the behaviour of $\Theta_{(0)}(\theta)$: given the value of $0 \le \theta \le \frac{\pi}{2}$, we use the red curve to determine the right hand side of \eqref{solu1}. We then map horizontally from the red curve to the blue curve to read off the value of 
	$\Theta_{(0)}$. We note that by symmetry
	\be
	\Theta_{(0)} \left  ( \frac{\pi}{4}  \right ) = \frac{\pi}{4}.
	\ee
	For $0 \le \theta < \frac{\pi}{4}$, $\Theta_{(0)} > \theta$ while for $\frac{\pi}{4} < \theta \le \frac{\pi}{2}$, $\Theta_{(0)} < \theta$. We can solve numerically for $\Theta_{(0)}(\theta)$; the plot for $K^{\frac{1}{2}} = 0.8$ is shown in Figure~\ref{fig:2}. Once $\Theta_{0)}(\theta)$ is determined, the other functions are determined using \eqref{R1c2}; the function $c_2 (\theta)$ is non-trivial for $K \neq 1$. 
	
	Thus, to summarise this section, matching a cosmic membrane metric on a constant null hypersurface with a Ricci flat metric with no deficits requires a relaxation of the asymptotically flat boundary conditions \eqref{afbc} to weaker boundary conditions of the form \eqref{weaker}. We will refer to Ricci flat metrics in Bondi gauge \eqref{bondi1} satisfying \eqref{weaker} as {\it asymptotically locally flat}. 
	
	\subsection{Cosmic membranes: alternative parameterisation}

	In the coordinate system of \eqref{cosmic5d2} the cosmic membrane preserves only a $U(1)$ symmetry, together with an additional inversion symmetry. To match such a metric, the required Bondi gauge parameterisation is
	\begin{eqnarray}
	ds^2 &=& - ({\cal U} e^{2 \beta} - r^2 W^2 e^{C_1}) du^2 - 2 e^{2 \beta} du dr - 2 r^2 W e^{C_1} du d \theta \label{bondi3} \nn\\
	&& + r^2 \left (e^{C_1} d \theta^2 +  \sin^2 \theta ( e^{C_2} d \chi^2 +  e^{- (C_1 + C_2)} \sin^2 \chi d \phi^2 ) \right  ), 
	\end{eqnarray}
	where the defining functions $({\cal U},W,\beta,C_1,C_2)$ can depend on $(u,r,\theta,\chi)$ but should be even functions of $\chi$, 
	to respect the inversion symmetry in \eqref{cosmic5d2}.
	
	\bigskip
	
	Now let us consider the matching between a Bondi gauge metric of the form \eqref{bondi3} and a cosmic membrane \eqref{cosmic5d2} 
	on a constant null time slice at infinity. The required coordinate maps are 
	\begin{eqnarray}
	U &=& U_{(0} (u, \theta, \chi) + \frac{1}{r} U_{(-1)}(u, \theta, \chi) + \cdots \\ 
	R &=& r R_{(1)} (\theta,\chi) + R_{(0)} (u, \theta, \chi) + \cdots\nn \\
	\Theta &=& \Theta_{(0)} (\theta,\chi) + \frac{1}{r} \Theta_{(-1)} (u, \theta,\chi) + \cdots \nn \\
	X &=& X_{(0)} (\theta,\chi) + \frac{1}{r} X_{(-1)} (u, \theta,\chi) + \cdots \nn \\
	\Phi &=& \phi \nn 
	\end{eqnarray}
	Again, the leading order terms in $(R,X,\Theta)$ are forced to be independent of $u$ to respect the asymptotic (local) flatness. 
	Matching on a null hypersurface then imposes four relations at leading order:
	\begin{eqnarray}
	( \theta \theta ): && \qquad R_{(1)}^2 \left ( (\partial_{\theta} \Theta_{(0)})^2 + \sin^2 \Theta_{(0)} (\partial_{\theta} X_{(0)})^2 \right ) 
	= e^{c_{1}(\theta,\chi)} \label{leading1a} \\
	(\theta \chi) :  && \qquad ( \partial_{\theta}  \Theta_{(0)} )( \partial_{\chi} \Theta_{(0)} ) + \sin^2 \Theta_{(0)} (\partial_{\theta} X_{(0)} )(\partial_{\chi} X_{(0)} ) = 0 \nonumber \\
	( \chi \chi): && \qquad R_{(1)}^2 \left ( (\partial_{\chi} \Theta_{(0)})^2 + \sin^2 \Theta_{(0)} (\partial_{\chi} X_{(0)})^2\right ) = e^{ c_2 (\theta,\chi)} \sin^2 \theta \nonumber \\
	(\phi \phi): && \qquad R_{(1)}^2 \sin^2 \Theta_{(0)} \sin^2 X_{(0)} = \frac{1}{K^2} e^{- c_1(\theta,\chi) - c_2 (\theta,\chi)} \sin^2 \theta \sin^2 \chi \nonumber
	\end{eqnarray}
	where we indicate which components of the induced metric are matched and we expand the defining functions $(C_1,C_2)$ as
	\be
	C_{i} (u,r,\theta,\chi) = c_{i}(\theta,\chi) + \frac{C_{i (- \lambda)}(u,\theta,\chi)}{r^{\lambda}} + \cdots
	\ee
	where the exponent $\lambda > 0$ will be determined by the Einstein equations. 
	
	The equations \eqref{leading1a} can clearly be solved by $X_{(0)} = \chi$, i.e. the coordinate transformations depend only on $\theta$ to leading order:
	\begin{eqnarray}
	( \theta \theta ): && \qquad R_{(1)}^2 (\partial_{\theta} \Theta_{(0)})^2 = e^{c_{1}(\theta)} \label{leading2a} \\
	( \chi \chi): && \qquad R_{(1)}^2 \sin^2 \Theta_{(0)}  = e^{ c_2 (\theta)} \sin^2 \theta \nonumber \\
	(\phi \phi): && \qquad R_{(1)}^2 \sin^2 \Theta_{(0)} = \frac{1}{K^2} e^{- c_1(\theta) - c_2 (\theta)} \sin^2 \theta  \nonumber
	\end{eqnarray}
	The last two equations are clearly not compatible for $K^2 \neq 1$ unless either one or both of $(c_1(\theta),c_2 (\theta))$ is non-zero: combining the last two equations we obtain
	\be 
	e^{c_1 + 2 c_2} = \frac{1}{K^2} \label{con1a}
	\ee
	However, as in the previous discussions, these equations are under-constrained: there are three equations for four functions, and thus one can fix a linear combination of $c_1$ and $c_2$ to be zero, provided that \eqref{con1a} is satisfied. The latter choice represents residual gauge freedom. 
	
	The equations \eqref{leading2a} clearly admit the solution
	\be
	\Theta_{(0)} = \theta; 
	\qquad R_{(1)} = \frac{1}{K^{\frac{1}{3}}}; \qquad
	e^{c_1} = e^{c_2} = \frac{1}{K^{\frac{2}{3}}}, 
	\ee
	i.e. an angle independent rescaling of the radius. This solution is trivial in the sense that the metric in coordinates $(r,\theta,\chi,\phi)$ still has a defect. 
	
	Combining the first two equations in \eqref{leading2a}, one obtains
	\be
	\frac{ \partial_{\theta} \Theta_{(0)}}{\sin \Theta_{(0)}} = \frac{e^{\frac{1}{2} (c_1 - c_2)}}{\sin \theta}. \label{redef1}
	\ee
	Suppose we fix a gauge in which
	\be
	e^{\frac{1}{2} (c_1 - c_2)} = \lambda, 
	\ee
	subject to the constraint \eqref{con1a}. Then \eqref{redef1} can be solved analogously to the angular equations of the previous sections. 
	
	\begin{figure}[tbp]
		\centering
		\includegraphics[width=0.5\linewidth]{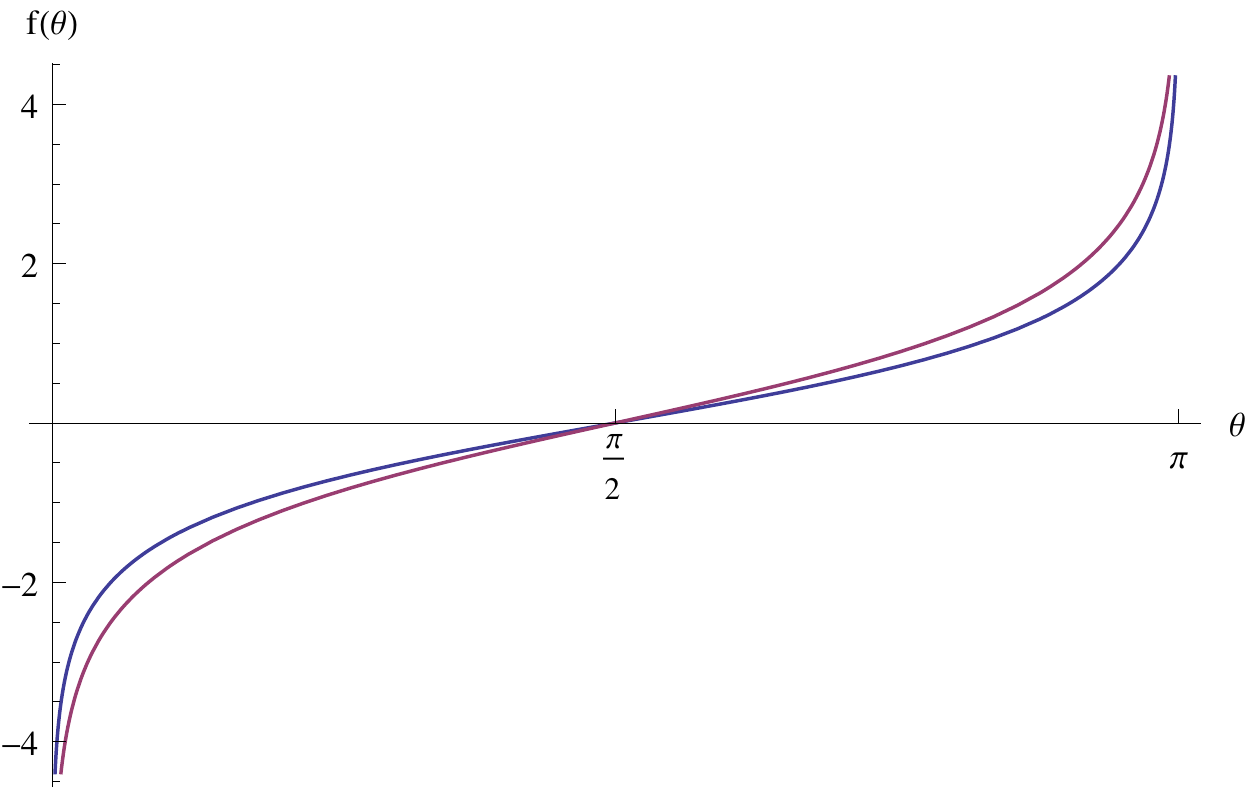}
		\caption{The red line plots $f(\theta)$ and the blue line plots $\lambda f(\theta)$ for $\lambda = 0.8$. For any value of $\lambda <  1$ the blue curve will lie closer to the horizontal axis than the red curve.}
		\label{fig:4}
	\end{figure}
	
\begin{figure}[tbp]
		\centering
		\includegraphics[width=0.5\linewidth]{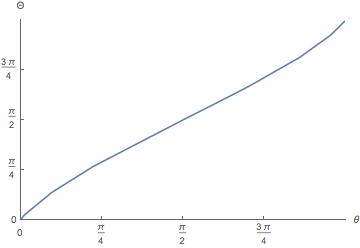}
		\caption{$\Theta_{(0)}(\theta)$ for $\lambda = 0.8$.}
		\label{fig:5}
	\end{figure}

	Let us define
	\be
	f(x) = \int \frac{dx}{\sin x} = \ln \left ( \tan \left ( \frac{x}{2} \right ) \right ). 
	\ee
	The integrated relation \eqref{redef1} can hence be expressed as 
	\be
	f(\Theta_{(0)}) = \lambda f (\theta). 
	\ee
	The function $f(\theta)$ is plotted over the range $(0,\pi)$ in Figure \ref{fig:4}. From the same plot we can see that if $\lambda < 1$ then the relation $\Theta_{(0)} (\theta)$ has a similar form to that in the previous section, see Figure~\ref{fig:5}: for $0 < \theta \le \pi/2$, $\Theta_{(0)} > \theta$ while for $\pi/2 \le \theta < \pi$, $\Theta_{(0)} < \theta$. 
	
	Note that for small $\theta$ 
	\be
	\Theta_{(0)} \approx 2 \left ( \frac{\theta}{2} \right )^{\lambda}
	\ee
	(with a corresponding expression for $\theta \sim \pi$). Furthermore, by symmetry, 
	\be
	\Theta{(0)} \left ( \frac{\pi}{2}\right ) = \frac{\pi}{2}. 
	\ee
	
	\subsection{Boundary conditions in $d$ dimensions and cosmic $(d-3)$ branes} \label{sec:Three-res}

	To match the cosmic brane metric to a non-singular Bondi gauge metric on a null hypersurface at infinity. we are forced to relax asymptotic flatness to asymptotic local flatness. The general Bondi gauge parameterisation (without imposing additional symmetries) of a spacetime in arbitrary dimension $d$ is 
	\be
	ds^2 = - {\cal U} e^{2 \beta} du^2 - 2 e^{2 \beta} du dr + r^2 h_{AB} (d \Theta^A + W^A du)(d \Theta^B + W^B du),
	\ee
	where the coordinates $\theta^A$ run from $A=1,\cdots (d-2)$. Here we have imposed the standard Bondi gauge conditions i.e. 
	\be
	g_{rr} = g_{rA} = 0 
	\ee
	and it is usual to impose the determinant condition 
	\be
	\partial_r \left ( {\rm det} (h_{AB} )  \right ) = 0. 
	\ee
	Asymptotically flat boundary conditions require that
	\be
	h_{AB} \rightarrow \gamma_{AB} + \frac{1}{r^{(d-2)/2}}h_{(d-2) AB} + \cdots 
	\ee
	with $\gamma_{AB}$ the metric on a unit $(d-2)$ sphere and the subleading term being associated with gravitational waves. 
	
	Such boundary conditions exclude cosmic $(d-3)$-branes passing through the celestial sphere. To allow for the latter, we need to relax the boundary conditions to asymptotically locally flat by setting
	\be
	h_{AB} \rightarrow h_{(0) AB} (\theta^C) + \cdots \label{alfc}
	\ee
	as $r \rightarrow \infty$. In the following sections we will impose such boundary conditions and consider the implications for the asymptotic structure in five dimensions. 
	
We should note that the boundary condition \eqref{alfc} is manifestly more general than that obtained from cosmic branes, for which
	$h_{(0)}$ is a spherical metric with distributional defects. As we discuss later, the main motivation for working with the more general boundary condition is holography.
	In (A)dS holography, the metric on the conformal boundary is allowed to be any non-degenerate metric, and it corresponds to the background metric for the dual quantum field theory. Even if one is only interested in the quantum field theory on a (conformally) flat background, one needs to allow for general perturbations of the boundary metric in order to compute correlation functions. If there is any holographic duality for asymptotically (locally) flat spacetimes, one would similarly expect that the spatial part of the boundary metric should be unrestricted. 
	
	If one takes a more conservative viewpoint and only wishes the boundary condition to be general enough to include distributional defects, one could regard the boundary condition \eqref{alfc} as encompassing all possibilities for distributional defects i.e. capturing different kinds of regularisations. One would then expect that asymptotic analysis of the field equations, combined with physical restrictions on allowed distributional curvature, will determine what additional restrictions should be placed on \eqref{alfc}.
		
	\section{Asymptotics and boundary conditions} \label{sec:four}
	
	In this section we will use the previous discussions of cosmic branes to postulate boundary conditions for asymptotically locally flat spacetimes in five dimensions. It is useful to start from the general definition of the Bondi gauge, following the original works \cite{Bondi1962,Sachs1962}. We define a function $u(x^A)$ which satisfies
	\be
	g^{\mu \nu} u_{, \mu} u_{, \nu} = 0
	\ee
	where we use $x^{\mu}$ to denote the five coordinates. Surfaces of constant $u$ are null hypersurfaces and the function $u$ can be interpreted as retarded time (Figure~\ref{fig:6}). As discussed in the previous section, we can write down a general form of the metric in coordinates $(u,r,x^A)$ as 
	\be
	ds^2 = - {\cal U} e^{2 \beta} du^2 - 2 e^{2 \beta} du dr + r^2 h_{AB} (d \Theta^A + W^A du)(d \Theta^B + W^B du). \label{bondi5}
	\ee
	Here we have imposed the standard Bondi gauge conditions i.e. 
	\be
	g_{rr} = g_{rA} = 0 
	\ee
	and it is also convenient to  impose the determinant condition 
	\be
	\partial_r \left ( {\rm det} (h_{AB} )  \right ) = 0. 
	\ee
	Using coordinates $(\theta,\psi, \phi)$ the three-dimensional metric $h_{AB}$ can be parameterised as
	\begin{equation}
	\begin{pmatrix}
	e^{C_1} & \cos \theta \sinh D_1 & \sin \theta \sinh D_2 \\
	\cos \theta \sinh D_1 & e^{C_2} \sin^2 \theta & \sin \theta \cos \theta \sinh D_3 \\
	\sin \theta \sinh D_2 & \sin \theta \cos \theta \sinh D_3 & e^{C_3} \cos^2 \theta 
	\end{pmatrix} \label{matr}
	\end{equation}
	where the determinant constraint implies that only five of the six functions are independent. 
	
	\bigskip
	
	The vacuum Einstein equations were analysed asymptotically in \cite{Tanabe:2009xb,Tanabe:2009va,Tanabe:2010rm}, under the assumption of asymptotic flatness, i.e. $h_{AB}$ asymptotes to the round metric on the unit three sphere and the subleading terms in the expansions arise from integration functions on solving the Einstein equations
	\be
	C_{i} \rightarrow \frac{C_{(\frac{3}{2}) i} (u,x^A)}{r^{\frac{3}{2}}} \qquad
	D_{i} \rightarrow \frac{D_{(\frac{3}{2}) i} (u,x^A)}{r^{\frac{3}{2}}}
	\ee
	as $r \rightarrow \infty$. Here $i = 1,2,3$ and the falloff behaviour relates to gravitational waves passing through null infinity. Without restricting to a specific choice of coordinates on the sphere, the expansion takes the form 
	\be
	h_{AB} = \gamma_{AB}  + \frac{C_{(\frac{3}{2}) AB}}{r^{\frac{3}{2}}} + \cdots
	\ee
	where $\gamma_{AB}$ is a round metric on a unit three sphere. The corresponding expansions of the other metric functions are  \cite{Tanabe:2009xb,Tanabe:2009va,Tanabe:2010rm}
	\begin{equation}
	{\cal U} = 1 + \frac{ {\cal U}_{(\frac{3}{2})}}{r^{\frac{3}{2}}} + \cdots \qquad
	\beta = \frac{\beta_{(3)}}{r^3} + \cdots \qquad
	W^A = \frac{W^A_{(\frac{5}{2})}}{r^{\frac{5}{2}}}  \cdots
	\end{equation}
	As discussed in previous sections, if we wish to impose weaker boundary conditions that would allow for cosmic branes, we should impose
	\be
	h_{AB} \rightarrow h_{(0) AB} (x^C)
	\ee
	as $r \rightarrow \infty$.
	
	\bigskip
	
	We can now analyse the asymptotic expansions of solutions to the vacuum Einstein equations with such boundary conditions. In Bondi gauge, the Einstein equations can as usual be split into main equations
	\be
	R_{rr} = R_{rA} = R_{AB} = 0
	\ee
	and supplementary (also called evolution) equations
	\be
	R_{uu} = R_{uA} = 0. 
	\ee
	The main equations determine the metric functions recursively, while the supplementary equations are automatically satisfied as a consequence of the Bianchi identities once the main equations are satisfied. 
	
	For computational simplicity we will continue to restrict to the case with $U(1)^2$ and reflection symmetry so that the functions $D_i$ defined in \eqref{matr} are zero. We can also eliminate $C_3$ using the determinant constraint i.e.
	\be
	C_3 = - (C_1 + C_2). 
	\ee 
	In this case there are five main equations $(R_{rr},R_{r\theta},R_{\theta \theta}, R_{\psi \psi}, R_{\phi \phi})$ and three supplementary 
	equations $(R_{uu},R_{ur},R_{u \theta})$. We will first write down the general form of these equations and then discuss asymptotic solutions. 
	
	\bigskip
	
	The $R_{rr}$ equation is
		\begin{empheq}[box=\fbox]{equation} \label{Rrr}
			R_{rr} = \frac{6}{ r} \beta_{,r} -  \frac{1}{2}  \left (  (C_{1,r})^2 + (C_{2,r})^2 + (C_{3,r})^2 \right ) = 0.
	 	\end{empheq} 
	Here and in the subsequent Einstein equations we denote partial derivatives with commas. 
	Clearly given $(C_1,C_2)$ this equation can be integrated to find $\beta$, with integration functions in both $\beta$ and $(C_1,C_2)$ left undetermined. 
	Note that this is exactly analogous to the well-known four-dimensional integration scheme: given the metric on the sphere, one can integrate to get $g_{ur}$. 
	
	Following the usual Bondi-Sachs integration scheme, we next use the $R_{r \theta}$ equation:
	\begin{empheq}[box=\fbox]{align}
	R_{r \theta} =&  \frac{1}{2 r^3} ( r^5 e^{C_1 - 2 \beta} W_{,r})_{,r} \label{Rrt} \nonumber\\
	& + \frac{1}{r} (3 \beta_{,\q} - r \beta_{,r\q}) + \frac{1}{2} ( (\cot \theta - 2 \tan \theta ) C_{1,r} + C_{1,r\q} ) \nonumber \\
	& - \frac{1}{4} ( 2 C_{1,\q} +C_{2,\q}) C_{1,r} - \frac{1}{4} \left( C_{1,\q} +  2 C_{2,\q}+ \frac{2}{\sin \theta \cos \theta}\right)C_{2,r}=0.  	\end{empheq} 
	Here and from now on we use the abbreviated notation $W \equiv W^{\theta}$. 
	Imposing $R_{r \theta} = 0$ allows us to integrate for $W$ in terms of $(C_i,\beta)$. 
	
	The three main equations in the sphere directions are as follows. The $R_{\q\q}$ equation is
	
\begin{empheq}[box=\fbox]{align} 
R_{\q\q}&= 2-2 (\b_{,\q})^2+\b_{,\q} C_{1,\q}-2 \b_{,\q\q} -\frac{1}{2} r^4(W_{,r})^2 e^{2 C_1-4 \b}\nonumber\\ &+C_{1,\q} (\cot \theta -\tan\theta)-\frac{1}{2}
	\csc\theta\sec \theta(C_{1,\q}-2C_{2,\q})\nonumber\\& -\frac{1}{2}\left((C_{1,\q})^2+(C_{2,\q})^2+ C_{1,\q}
	C_{2,q}- C_{1,\q\q}\right) \nonumber\\
	& +e^{C_1-2\b}(4 r W_{,\q} +r W 
	(\cot\theta-\tan\theta)-2 \mathcal{U}) \nonumber\\&
	+e^{C_1-2 \b} \left(\frac{1}{2} r^2 C_{1,\q}W_{,r}-\frac{3}{2} r \mathcal{U} C_{1,r}+\frac{3}{2} r W
	C_{1,\q}-r U_{,r}\right)\nonumber \\ &
	+\frac{1}{2} r^2 e^{C_1-2 \b} \left(C_{1,r}
	\left(W_{,\q}-\mathcal{U}_{,r}\right)+W C_{1,r} (\cot
	\theta -\tan \theta )\right)\nonumber \\ &
	+r^2 e^{C_1-2 \b} \left(\frac{3 C_{1,u}}{2 r}-\frac{1}{2} \mathcal{U}
	C_{1,rr}+W
	C_{1,r\q}+C_{1,ur}+W_{,r\q}\right)=0
	\end{empheq}
	The $R_{\phi\phi}$ equation is
	\begin{empheq} [box=\fbox]{align} 
R_{\phi\phi}& =e^{2 \b}\left(-2\b_{,\q} C_{2,\q}+ C_{2,\q} \tan\theta+C_{1,\q} C_{2,\q}- C_{2,\q\q}\right)\nonumber\\
	&+e^{2 \b}(4+\cot \theta(
	2C_{1,\q}-4\b_{,\q}-C_{2,\q}
	-2 e^{C_1} r^2 W_{,r})\nonumber\\&+r^2e^{C_1} \left(C_{2,\q} W_{,r}+C_{2,r} W_{,\q}+2 C_{2,ur} -C_{2,r} \mathcal{U}_{,r}- \mathcal{U}
	C_{2,rr}\right)\nonumber\\&+re^{C_1}\left(3 C_{2,u}+3  W C_{2,\q}-3 \mathcal{U} C_{2,r}-2\mathcal{U}_{,r}+2 W_{,\q}\right)\nonumber \\ &
	-4 e^{C_1} \mathcal{U} +e^{C_1} r W ((5+rC_{2,r}) (\cot\q-\tan\q)+ 2rC_{2,r\q})\nonumber \\ &
	+3e^{C_1} r W\sec \theta \csc \theta  =0\,.
	\end{empheq}
	The $R_{\psi\psi}$ equation is, applying $C_{3}=-(C_1+C_2)$ to simplify, 
	\begin{empheq}[box=\fbox]{align}
	R_{\psi\psi}&= \sin \theta  \left(4 e^{2 \b} \b_{,\q}-3
	e^{2 \b} C_{1,\q}-e^{2 \b}
	C_{2,\q}-2 e^{C_1} r^2
	W_{,r}\right)\nonumber \\ 
	&+\cos \theta C_{1,\q} \left(2
	e^{2 \b} \b_{,\q}-e^{2 \b}
	C_{2,\q}-e^{C_1} r^2
	W_{,r}+e^{2 \b} \cot\theta
	\right) \nonumber \\ 
	& +e^{2 \b}\cos \theta  \left(2  \b_{,\q}
	C_{2,\q}+ C_{2,\q}
	\cot\theta-
	\left(C_{1,\q}\right)^2-
	C_{3,\q\q}+4 \right)\\ 
	&+e^{C_1}\cos \theta  \left(-r^2
	C_{2,\q} W_{,r}-2  r
	\mathcal{U}_{,r}+ r W_{,\q} \left(r
	C_{3,r}+2\right)\right)\nonumber \\ 
	&+e^{C_1} r W\csc\theta \left(r C_{3,r\q}
	\sin 2 \theta+r C_{3,r} \cos2 \theta
	+5 \cos2 \theta -3\right)) \nonumber \\ 
	&+e^{C_1} \cos\theta \left(r \left(2 rC_{3,ur}+3 C_{3,u}\right)-\mathcal{U} \left(r^2 C_{3,rr}+3 r C_{3,r}+4\right)\right)\nonumber\\&
	+re^{C_1}\cos
	\theta (3 W  C_{3,\q} - r
	C_{3,r} \mathcal{U}_{,r})=0. \nonumber
	\end{empheq}
	Combining these equations to form the trace along the sphere, i.e $g^{AB} R_{AB} = 0$, one obtains an equation that determines ${\cal U}$ from the previously determined $(\beta,W)$ and $C_i$
	\begin{empheq}[box=\fbox]{align}
	g^{AB}R_{AB} &=-\frac{e^{-C_1}}{2r^2}\sec\q\csc\q( C_{1,\q}+2C_{2,\q}) \label{Rtrace}\nonumber \\
	& +\frac{e^{-2\beta-C_1}}{r^2}(\cot\theta-\tan\theta)\left(e^{C_1}(6rW+r^2W_{,r})-2 e^{2\beta}\beta_{,\q}+\frac{5}{2}e^{2\beta} C_{1,\q}\right) \nonumber\\
	&+\frac{e^{-C_1}}{2r^2}\left(12-(2\b_{,\q})^2+4\beta_{,\q} C_{1,\q}-2( C_{1,\q})^2- C_{1,\q} C_{2,\q}-( C_{2,\q})^2\right)\nonumber\\&+\frac{e^{-C_1}}{2r^2} \left(-4\beta_{,\q\q}+2C_{1,\q\q}\right)-\frac{r^2}{2}e^{-4\beta+C_1}(W_{,r})^2\nonumber\\
	&-3\frac{e^{-2\beta}}{r^2}\left((2+r\partial_r)\mathcal{U}-r(2+\frac{r}{3}\partial_r)W_{,\q}\right)=0\,.
	\end{empheq}
Having solved this equation, the remaining main equations then determine the $u$ evolution of the metric functions along the sphere $C_i$ from the original data $C_i$ and the determined functions $(\beta,W,{\cal U})$. 
	
	The supplementary equations are 
	\begin{align}
	R_{uu}&=+\frac{1}{2}r^4 e^{2 C_1-4 \b} W^2 \left(W_{,r}\right)^2+r^2e^{C_1-2 \b}W^2
	\left(\frac{1}{2} \mathcal{U}_{,r} C_1{}_{,r}-2 W_{,r\q}+\frac{1}{2} \mathcal{U}
	C_1{}_{,rr}\right)\nonumber\\&+r^2e^{C_1-2 \b}W
	\left(-C_1{}_{,ur} W-2 W_{,\q} W_{,r}-2 \mathcal{U}
	\b_{,r} W_{,r}-C_1{}_{,r,\q} W^2+2 \b_{,\q} W_{,r} W\right) \nonumber \\
	&+r^2e^{C_1-2 \b}
	\left(-\frac{3}{2}
	C_1{}_{,\q} W_{,r} W^2-\frac{1}{2} W_{,\q} C_1{}_{,r}
	W^2+\mathcal{U} W_{,r} C_1{}_{,r} W+\mathcal{U} W_{,rr} W-W_{,ur}
	W\right) \nonumber  \\
	& 
	+r^2e^{C_1-2 \b}
	\left(+2
	W_{,r} \b_{,u} W-W_{,r} C_1{}_{,u} W+\frac{1}{2} \mathcal{U} \left(W_{,r}\right)^2\right) \nonumber \\
	& +r e^{C_1-2 \b}
	\left(-\frac{3}{2} C_1{}_{,\q} W^3-4 W_{,\q} W^2+\mathcal{U}_{,r}
	W^2+\frac{3}{2} \mathcal{U} C_1{}_{,r} W^2-\frac{3}{2} C_1{}_{,u} W^2+5 \mathcal{U}
	W_{,r} W\right) \nonumber \\
	& +2W^2(e^{C_1-2 \b} \mathcal{U}+
	\left(\b_{,\q}\right)^2)-W^2 \b_{,\q} C_1{}_{,\q}+\left(W_{,\q}\right)^2+\frac{1}{2}
	\left(C_1{}_{,u}\right){}^2+\frac{1}{2} \left(C_2{}_{,u}\right){}^2-2
	W \b_{,\q} W_{,\q}\nonumber \\
	&  +W W_{,\q}C_1{}_{,\q}+W W_{,\q\q}+\frac{1}{2} W^2 C_1{}_{,\q\q}+2 W \mathcal{U}_{,\q}\b_{,r}+\mathcal{U} W_{,\q} \b_{,r}+\frac{1}{2} W_{,\q} \mathcal{U}_{,r}-\mathcal{U}
	\b_{,r} \mathcal{U}_{,r}\nonumber \\ 
	&+\mathcal{U} \b_{,\q} W_{,r}-\frac{1}{2} \mathcal{U}_{,\q}
	W_{,r}-W \mathcal{U}_{,\q} C_1{}_{,r}+2 \mathcal{U} W \b_{,r\q}+W \mathcal{U}_{,r\q}-\mathcal{U}^2
	\b_{,rr}-\frac{1}{2} \mathcal{U} \mathcal{U}_{,rr}-2 W_{,\q} \b_{,u}\nonumber \\
	&
	+W_{,\q}
	C_1{}_{,u}+\frac{1}{2} C_1{}_{,u} C_2{}_{,u}-2 W
	\b_{,u\q}+W_{,u\q}+W C_1{}_{,u\q}+2 \mathcal{U} \b_{,ur} \nonumber \\
	&+\frac{e^{-4 \b-C_1}}{2 r^2}(\cot \theta-\tan\theta ) \left(e^{6 \b} \left(2 U \b_{,\q}+U_{,\q}\right)+r^3 W^2 e^{2
		\left(\b+C_1\right)} \left(W \left(r C_1{}_{,r}+2\right)+2 r W_{,r}\right)\right)\nonumber\\
	& -\frac{1}{2} (\cot \theta-\tan
	\theta ) \left(2 U W \b_{,r}-4 W \b_{,u}+W^2 C_1{}_{,\q}+2 W C_1{}_{,u}+W U_{,r}+2 W W_{,\q}+2
	W_{,u}\right)\nonumber \\ 
	&
	+\frac{1}{r}(-3 \b_{,r}
	\mathcal{U}^2+3 W \b_{,\q} \mathcal{U}-\frac{3}{2} \mathcal{U}_{,r} \mathcal{U}+3 \b_{,u} \mathcal{U}-\frac{1}{2} W
	\mathcal{U}_{,\q}-\frac{3}{2} \mathcal{U}_{,u}) \nonumber \\
	&+\frac{e^{2 \b-C_1}}{r^2} \left(-2 \mathcal{U}
	\left(\b_{,\q}\right)^2-\mathcal{U}_{,\q} \b_{,\q}+\mathcal{U} C_1{}_{,\q}
	\b_{,\q}+\frac{1}{2} \mathcal{U}_{,\q} C_1{}_{,\q}-\mathcal{U} \b_{,\q\q}-\frac{1}{2}
	\mathcal{U}_{,\q\q}\right) = 0\,,\quad
	\end{align}
	and
	\begin{align}
	R_{u\theta}&= \frac{1}{4} e^{-2 \b} (\cot \q-\tan \q) \left(W \left(4 e^{2 \b}
	\b_{,\q}-2 e^{C_1} r^2 W_{,r}\right)+3 e^{2 \b} C_1{}_{,u}-2
	e^{C_1} r W^2 \left(r C_1{}_{,r}+2\right)\right) \nonumber \\ 
	& -\frac{1}{4}
	\left(C_1{}_{,u}+2 C_2{}_{,u}\right) \csc \q \sec\q-2 r^3 e^{2 \b+C_1} \left(W_{,r} \left(C_1{}_{,u}-2 \b_{,u}\right)+2
	W_{,\q} W_{,r}+W_{,ur}\right) \nonumber\\
	&
	+\frac{\mathcal{U}_{,\q}}{4 r}\left(4 r
	\b_{,r}-2 r C_1{}_{,r}+2\right)+W
	\left(-\b_{,\q} C_1{}_{,\q}+2
	\left(\b_{,\q}\right)^2+\b_{,\q\q}\right)+\frac{1}{2} r^4 W
	\left(W_{,r}\right)^2 e^{2 C_1-4 \b} \nonumber \\ 
	&+\frac{1}{4} \left(-4
	\b_{,u\q}-C_2{}_{,\q} C_1{}_{,u}-2 C_2{}_{,\q}
	C_2{}_{,u}-C_1{}_{,\q} \left(2
	C_1{}_{,u}+C_2{}_{,u}\right)+2 C_{1,u\q}+2 \mathcal{U}_{,r\q}\right) =0\,.\quad
	\end{align}
	The first supplementary equations gives the $u$-evolution equation for the free data in the $\mathcal{U}$ expansion. 
	The second gives the $u$ evolution of the free data in $W$.
	
	The final supplementary equation (sometimes also called the trivial equation) is:
	\begin{eqnarray}
	R_{ur}&=& -\frac{e^{-C_1}}{2 r^2} (\cot\theta-\tan \theta)
	\left(-2 e^{2 \b} \b_{,\q}+e^{C_1} r^2
	W_{,r}+e^{C_1} r W \left(r
	C_1{}_{,r}+2\right)\right) \nonumber \\ 
	&& +r^2e^{C_1-2
		\b}
	\left(W \b_{,r} W_{,r} -\frac{1}{2} \left(W_{,r}\right)^2-\frac{1}{2} W C_1{}_{,r}
	W_{,r}-\frac{1}{2} W W_{,rr}-\frac{5}{2} r W W_{,r}\right) \nonumber \\
	&& +\frac{2}{r^2}e^{2
		\b-C_1}(
	\left(\b_{,\q}\right)^2 -\b_{,\q} C_1{}_{,\q} +\b_{,\q,\q})+\frac{3}{r} (\mathcal{U}
	\b_{,r}-3 W \b_{,\q}+\frac{3}{2}
	\mathcal{U}_{,r}-W_{,\q}) \nonumber \\ 
	&& +\b_{,r}
	\mathcal{U}_{,r}+\mathcal{U} \b_{,rr}-\b_{,\q}
	W_{,r}-W \b_{,r,\q}-2
	\b_{,u,r} -\frac{1}{2} C_1{}_{,r}
	W_{,\q}-\frac{1}{2} W
	C_1{}_{,r\q} \nonumber \\ 
	&& -\frac{1}{2} C_1{}_{,r}
	C_1{}_{,u}-\frac{1}{4} C_2{}_{,r}
	C_1{}_{,u}-\frac{1}{4} C_1{}_{,r}
	C_2{}_{,u}-\frac{1}{2} C_2{}_{,r}
	C_2{}_{,u}+\frac{1}{2}
	\mathcal{U}_{,r,r}-\frac{1}{2} W_{,r\q} =0\,.\qquad\quad
	\end{eqnarray}
	This is automatically satisfied at each order as a consequence of Bianchi identities once the main equations are satisfied. This equation therefore does not give any new information. If we do not check Bianchi identities, then the vanishing of this equation at each order provides a check on our solution.
	
	\subsection{Asymptotic analysis}\label{Asym}
	
	Having collected together the Einstein equations, let us consider asymptotic solutions of these equations. The equations in Bondi gauge are nested, and thus should be solved in the order in which they are presented above, beginning with \eqref{Rrr}. 
	
	\begin{figure}[tbp]
		\centering
		\includegraphics[width=0.4\linewidth]{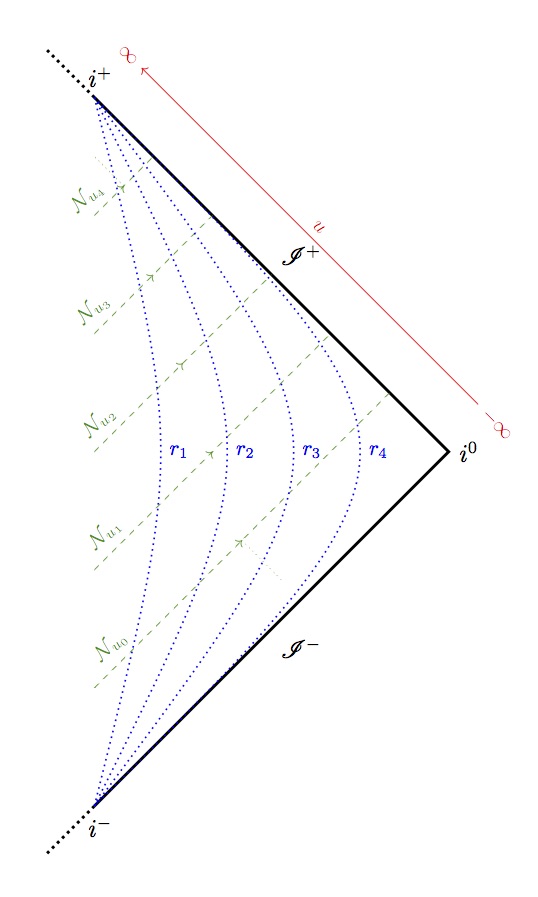}
		\caption{Penrose diagram indicating hypersurfaces of constant $u$ and $r$.}
		\label{fig:6}
	\end{figure}

	As we described above, in the Bondi integration scheme we prescribe data for $C_i$ on a null hypersurface, say $\mathcal{N}_{u0}$ in Figure~\ref{fig:6}, recursively determine the other metric coefficients using the main equations and then determine the null evolution of $C_i$ using the final null equation. Thus we should impose boundary conditions for $C_i$ as $r \rightarrow \infty$, and examine their consequences for the nested integration. Following the discussions of the previous section, we impose the boundary condition that $
	C_{i} \rightarrow C_{i(0)} (\theta)$
	as $r \rightarrow \infty$. The $u$ independence was established in the previous section but we will understand further below why the final main equation requires $u$ independence of the defining data on the celestial sphere. The corresponding asymptotic expansion of $C_i$ is therefore
	\be
	C_i = C_{i(0)} (\theta) + \cdots 
	\ee 
	where the ellipses denote subleading terms in the radial expansion. The structure of these subleading terms will be determined below by the field equations i.e. we do not make any assumptions a priori for the form of the expansion. 
	
	We can trivially rearrange the first main equation \eqref{Rrr} to obtain 
	\boxedeq{Rrr2}
{	\beta_{,r} =  \frac{r}{6} \left ( C_{1,r}^2 + C_{1,r} C_{2,r} + C_{2,r}^2
	\right  ).}
	Integrating this equation the leading contribution to $\beta$ is an integration function
	\be
	\beta = \beta_{(0)} (\theta,u) + \cdots 
	\ee
	Note that this is clearly the only integration function from this equation to all orders in the radial expansion. According to the standard analysis in four and higher dimensions this integration function is set to zero to satisfy the boundary conditions.
	
	
	\bigskip
	
	Moving now to the \eqref{Rrt} equation, we can write this in the form
	\boxedeq{beta2alpha}
{	\frac{1}{2 r^3}  ( r^5 e^{C_1 - 2 \beta} W_{,r} )_{,r} = {\cal G} (C_1,C_2,\beta)}
	with ${\cal G}$ as given in \eqref{Rrt}. Integrating for $W$ we obtain
	\be
	W = W_{(0)} + \frac{W_{(1)}[C_{i(0)},\b_{(0)}]}{r} + \cdots
	\ee
	where $[\cdots]$ denotes the functional dependence of the coefficients. $W_{(0)}$ is again an integration function and the coefficient $W_{(1)}$ is completely determined by the $1/r$ terms in the $R_{r \theta}$ equation:
	\be
	W_{(1)} = 2e^{2 \beta_{(0)} - C_{1(0)}} \partial_{\theta} \beta_{(0)}.
	\ee
	The only way to satisfy the Einstein equation at this order is either to allow for $W_{(1)}$ or to fix $\beta_{(0)}$ to be independent of $\q$. However, we will see that the function $W_{(0)}$ must actually be set to zero, as a consequence of the next equation. 
	%
	
	\bigskip
	
	Next we consider the trace of the main equations along the sphere \eqref{Rtrace}. One can write this in the form
	\boxedeq{Rtrace2}
{	3 \frac{e^{-2 \beta}}{r^2} ( 2 + r \partial_r) {\cal U} = {\cal F} (C_1,C_2,\beta,W), }
	where the leading contribution to the functional ${\cal F}$ is at order\footnote{Note that $(\partial_{\q}+(\cot\q-\tan\q))W=D_{\theta}W^{\q}$ where $D$ is the covariant derivative.} $1/r$ 
	\begin{equation}
	\mathcal{F}=6e^{2\b_{(0)}}=6e^{2\b_{(0)}}\left(\partial_{\q}+(\cot\q-\tan\q)\right)W_{(0)}.
	\end{equation}
	Then \eqref{Rtrace2} implies that 
	\be
	{\cal U} = r {\cal U}_{(-1)}[W_{(0)}] + {\cal U}_{(0)}[C_{i(0)},\b_{(0)}]+ \cdots 
	\ee
	where 
	\be
	{\cal U}_{(-1)} = \frac{2}{3}( \partial_{\theta} + (\cot \theta - \tan \theta)) W_{(0)}. 
	\ee
	However, this solution implies that 
	\be
	g_{uu} = r^2 e^{C_1} W^2 - {\cal U} e^{2 \beta} \rightarrow r^2 e^{C_{1(0)}} W_{(0)}^2 
	\ee
	as $r \rightarrow \infty$ i.e. $\partial_{u}$ is spacelike rather than null or timelike. The requirement that $\partial_u$ is not spacelike at infinity thus fixes
	$W_{(0)} = 0$. 
	
	The physical interpretation of non-zero $W_{(0)}$ can be understood using the example of Minkowski spacetime in four dimensions. Starting from
	\be
	ds^2 = - du^2 - 2 du dr + r^2 (d \theta^2 + \sin^2 \theta d \phi^2)
	\ee
	we can change coordinates to 
	\be
	d \phi = d \tilde{\phi} + \frac{du}{\Omega} 
	\ee
	(where $\Omega$ characterises the angular velocity) so that
	\be
	ds^2 = - du^2 - 2 du dr + r^2 \left (d \theta^2 + \sin^2 \theta \left (d \tilde{\phi} + \frac{du}{\Omega} \right )^2  \right )
	\ee
	i.e. comparing with \eqref{bondi5} $W^{\tilde{\phi}}_{(0)} \neq 0$. Thus, physically, a non-zero $W_{(0)}$ can be interpreted 
	as using a rotating frame at null infinity. 
	
	\bigskip
	
	Setting $W_{(0)} = 0$, the leading contribution to the function ${\cal F}$ in \eqref{Rtrace2} is at order $1/r^2$:
	\begin{eqnarray}
	{\cal F}_{(2)} &=& - \frac{e^{-C_{1(0)}}} {2 \sin\theta\cos\theta}( \partial_\theta C_{1(0)}+ 2 \partial_\theta C_{2 (0)}) 
	- e^{-C_{1(0)}}  (\cot\theta-\tan\theta)(2 \partial_\theta\beta-\frac{5}{2}\partial_\theta C_{1(0)}) \nonumber\\
	&& +e^{-C_{1(0)}}
	(6 -2 (\partial_\theta\beta_{(0)})^2 - 2 \partial_\theta^2\beta_{(0)} 
	+2 \partial_\theta\beta _{(0)} \partial_\theta C_{1(0)} ) \\
	&& - \frac{e^{-C_{1(0)}}} {2 } (
	2(\partial_\theta C_{1(0)})^2+\partial_\theta C_{1(0)}\partial_\theta C_{2(0)} +(\partial_\theta C_{2(0)})^2-4\partial_\theta^2 C_{1(0)}) \nonumber 
	\end{eqnarray}
	and thus integrating \eqref{Rtrace2} we obtain
	\be
	{\cal U}_{(0)} = \frac{1}{6} e^{2 \beta_{(0)}} {\cal F}_{(2)}.  \label{uo}
	\ee
	For $\partial_{u}$ to be non-spacelike as $r \rightarrow \infty$ we require that
	\be
	e^{C_{1(0)}} W_{(1)}^2 - e^{2 \beta_{(0)}} {\cal U}_{(0)} \le 0 
	\ee
	i.e.
	\be
	{\cal F}_{(2)} \ge 24 e^{-C_{1(0)}} (\partial_\theta\beta_{(0)})^2 
	\ee
	so
	\begin{eqnarray}
	&& 6 - 26 (\partial_\theta\beta_{(0)})^2 - 2 \partial_\theta^2\beta_{(0)} 
	+ 2 \partial_\theta\beta _{(0)} \partial_\theta C_{1(0)}  \\
	&& - \frac{1}{2 \sin\theta\cos\theta}( \partial_\theta C_{1(0)}+ 2 \partial_\theta C_{2 (0)}) 
	-  (\cot\theta-\tan\theta)( 2 \partial_\theta\beta-\frac{5}{2}\partial_\theta C_{1(0)}) \nonumber \\
	&& - \frac{1}{2} 
	\left (2 (\partial_\theta C_{1(0)})^2+\partial_\theta C_{1(0)}\partial_\theta C_{2(0)} +(\partial_\theta C_{2(0)})^2-4\partial_\theta^2 C_{1(0)} \right ) \ge 0  \nonumber 
	\end{eqnarray} 
	This is a non-trivial constraint. In the case of cosmic membranes discussed previously the functions $(C_{i(0)}, \beta_{(0)})$ are proportional to the membrane tension; provided that the tension is much less than one, the leading term in this expression will be the first one and the constraint be satisfied. In other words, for a cosmic membrane, 
	\be
	{\cal U}_{(0)} \approx 1,
	\ee
	up to corrections of order the membrane tension.
	
	\bigskip
	
	The remaining Einstein equations do not place further constraints on this defining data. The Einstein equations along the sphere can be expressed in the form:
	\begin{empheq}[box=\fbox]{align} 
	R_{\theta\theta}=0\Leftrightarrow (3 r + 2 r^2 \partial_r) \partial_u C_1 =& {\cal H}_{1} (C_i,\beta,W, {\cal U});  \label{Rsphere} \\
	R_{\phi\phi}=0\Leftrightarrow (3 r + 2 r^2 \partial_r) \partial_u C_2 =& {\cal H}_{2} (C_i,\beta,W, {\cal U}); \nonumber \\
	R_{\psi\psi}=0\Leftrightarrow (3 r + 2 r^2 \partial_r) \partial_u C_3 =& {\cal H}_{3} (C_i,\beta,W, {\cal U}), \nonumber
	\end{empheq}
	and these determine the $u$ evolution of the functions $C_i$.  Here the functionals ${\cal H}_{i}$ depend on the functions
	$(C_i,\beta,W,{\cal U})$ and their $(r,\theta)$ derivatives, but not on $u$ derivatives. The three equations are not independent: $C_3 = C_{1} + C_{2}$. 
	
	Asymptotically, the leading contributions to ${\cal H}^i$ are of order one, thus determining that there are terms at order $1/r$ in the $C_i$ expansions
	\be\label{cexp}
	C_i = C_{i(0)}(\theta)  + \frac{C_{i(1)}(u,\theta)}{r} + \cdots  
	\ee 
	The equations \eqref{Rsphere} can immediately be integrated at leading order to give
	\be
	C_{i (1)} = \int {\cal H}_{i} (u,\theta) du. 
	\ee
	where ${\cal H}_i$ are
	\begin{align}
	{\cal H}_1&=\frac{1}{3} e^{2 \b_{(0)}-C_{1(0)}} \left((C_{1(0),\q})^2+2 C_{2(0),\q}
	C_{1(0),\q}+2(C_{2(0),\q})^2-C_{1(0),\q\q}\right)\nonumber\\&
	+\frac{2}{3} e^{2 \b_{(0)}-C_{1(0)}}\left(
	C_{1(0),\q}  \b_{(0),\q}-8
	( \b_{(0),\q})^2-4 \b_{(0),\q\q}\right)\nonumber\\&+\frac{1}{3} e^{2 \b_{(0)}-C_{1(0)}}(\tan \q-\cot \q) \left(C_{1(0),\q}-4 \b_{(0),\q}\right)\nonumber\\&
	+\frac{2}{3} e^{2 \b_{(0)}-C_{1(0)}}\csc\q \sec\q \left( C_{1(0),\q}+2
	C_{2(0),\q}\right)
	\end{align}
	and
	\begin{align}
	{\cal H}_2&= \frac{1}{3} e^{2 \b_{(0)}-C_{1(0)}} \left(-4\b_{(0),\q} C_{1(0),\q}-6 \b_{(0),\q}
	C_{2(0),\q}\right)\nonumber \\& + \frac{1}{3} e^{2 \b_{(0)}-C_{1(0)}}\left( +8
	(\b_{(0),\q})^2+4 \b_{(0),\q\q}-2
	C_{1(0),\q}^2\right)\nonumber\\&-\frac{2}{3} e^{2 \b_{(0)}-C_{1(0)}}(\tan\q+\cot\q)C_{1(0),\q}\nonumber\\&+\frac{1}{3}e^{2 \b_{(0)}-C_{1(0)}}(\cot\q-5 \tan\q) C_{2(0),\q} \nonumber\\&-\frac{4}{3}e^{2 \b_{(0)}-C_{1(0)}}( \tan\q+ 2\cot\q)
	\b_{(0),\q}\nonumber\\&
	-\frac{1}{3} e^{2 \b_{(0)}-C_{1(0)}}\left((C_{2(0),\q})^2+4 C_{1(0),\q} C_{2(0),\q}\right)\nonumber\\&
	+\frac{1}{3} e^{2 \b_{(0)}-C_{1(0)}}\left(2 C_{1(0),\q\q}3
	C_{2(0),\q\q}\right)
	\end{align}
Note that  ${\cal H}_3$ is the sum of ${\cal H}_1$ and ${\cal H}_2$. 
		
The equations \eqref{Rsphere} demonstrate why the defining data $C_{i(0)}$ should be independent of $u$ as $r \rightarrow \infty$: these equations cannot be solved self-consistently if $C_{i(0)}$ depends on $u$, without inducing an infinite series of terms in $C_i$ that scale as positive powers of $r$, hence breaking the notion of local flatness. 
	
	\bigskip
	
	Thus, in summary, the leading terms in the asymptotic expansions are
	\begin{eqnarray}
	C_i &=& C_{i(0)} + \cdots \qquad 
	\beta = \beta_{(0)} + \cdots  \\
	W &=& \frac{W_{(1)}}{r} + \cdots \qquad 
	{\cal U} = {\cal U}_{(0)} + \cdots \nonumber
	\end{eqnarray}
	where $(C_{i(0)}(\theta),\beta_{(0)}(u,\theta))$ are the independent data and $(W_{(1)},{\cal U}_{(0)})$ are determined from this data. 
	Note that if $W_{(1)}$ is non-zero then $g_{u \theta} \sim r$. By setting $\partial_{\theta} \beta_{(0)} = 0$, one can set $W_{(1)} = 0$. If 
	$\beta_{(0)}$ is a function only of $u$, one can then use reparameterisation freedom of the retarded time coordinate to fix $\beta_{(0)} = 0$.  In this case $C_{i(0)}$ is the only remaining non-trivial data, with ${\cal U}_{(0)}$ determined from this data via \eqref{uo}. 
	
	\bigskip
	
	One can use this behaviour to write down the asymptotics of the metric in the general case:
	\begin{eqnarray}\label{leading_terms}
	g_{AB}&=&r^2h_{AB}=r^2h_{(0)AB} + {\cal O}(r)  \\
	g_{uu}&=&(-{\cal U}e^{2\b}+r^2h_{AB}W^AW^B) =  -{\cal U}_{(0)}e^{2\b_{(0)}}+h_{(0)AB}W_{(1)}^AW_{(1)}^B+ O(r^{-1}) \nonumber \\
	g_{ur}&=& -e^{2\b}=-e^{2\b_0} + {\cal O}(r^{-2}) \nonumber\\
	g_{uA}&=& g_{AB}W^B = r h_{(0)AB}W_{(1)}^B + {\cal O}(r^0), \nonumber
	\end{eqnarray}
	where the orders of the subleading terms follow from the next to leading contributions to the Einstein equations. Note that if one imposes the additional constraint that
	\be
	\beta_{(0), A} = 0
	\ee
	then $W_{(1)}^A = 0$ and $g_{uA}$ is order $r^0$, as in the four-dimensional case. 

	\section{Asymptotic expansion to all orders} \label{sec:five}
	
In the previous section we established the leading asymptotics for the metric components, given the generalised boundary condition for the metric on the celestial sphere. In this section we will establish the asymptotic expansion for the metric in this context. It is useful to first review the structure of the expansion of an {\it asymptotically flat} vacuum metric, analysed in detail in \cite{Tanabe:2009xb,Tanabe:2009va,Tanabe:2010rm}. For direct comparison with the results above, we restrict to $U(1)^2$ symmetry and inversion symmetry, i.e. we set $D_i = 0$. The expansions for the five metric functions are then
	\begin{eqnarray}
	C_i &=& \frac{ \textcolor{blue}{ C_{i ( \frac{3}{2} )} }} {r^{\frac{3}{2}}} +  \cdots \\
	\beta &=&  \frac{\beta_{(3)}}{r^3} + \cdots \nonumber \\
	W &=& \frac{W_{( \frac{5}{2} )}}{r^{\frac{5}{2}}} + \frac{W_{(3 )}}{r^{3}} + \frac{W_{( \frac{7}{2} )}}{r^{\frac{7}{2}}} + \frac{ \textcolor{red}{ W_{(4)}}}{r^4} + \cdots \nonumber \\
	{\cal U} &=& 1 + \frac{ {\cal U}_{ ( \frac{3}{2} )}}{r^{\frac{3}{2}}} + \frac{ \textcolor{red}{{\cal U}_{(2)}}}{r^2} + \cdots \nonumber 
	\end{eqnarray}
	Here we have highlighted in colour the defining data for the asymptotic expansion; all other expansion coefficients can be expressed in terms of this data and its derivatives, with explicit expressions given in \cite{Tanabe:2009xb,Tanabe:2009va,Tanabe:2010rm}.The integration functions $\textcolor{blue}{C_{i ( \frac{3}{2} )}}$ are associated with gravitational wave degrees of freedom and their $u$-evolution is not fixed by Einstein equations and has to be prescribed to fully determine the solution. The integration functions highlighted in red are associated with conserved charges; in particular, $\textcolor{red}{{\cal U}_{(2)}}$ is associated with the mass of the spacetime. (Note that the assumed symmetries set the angular momentum charges associated with rotations in the $\phi$ and $\psi$ directions to zero.)

\bigskip
	
We now turn to the asymptotic expansions of {\it asymptotically locally flat} vacuum metrics. The expansions for the five metric functions are 
	\begin{eqnarray}\label{ALFexpansion}
	C_i &=& \textcolor{ao(english)} {C_{i (0)} } + \frac{C_{i (1)}}{r} + \frac{ \textcolor{blue}{ C_{i ( \frac{3}{2} )} }} {r^{\frac{3}{2}}} +  \cdots \\
	\beta &=& \textcolor{ao(english)} { \beta_{(0)}} + \frac{\beta_{(2)}}{r^2} +  \frac{\beta_{(\frac{5}{2})}}{r^{\frac{5}{2}}} + \frac{\beta_{(3)}}{r^3} + \cdots \nonumber \\
	W &=& \frac{W_{(1)}}{r} + \frac{W_{(2)}}{r^2} + \frac{W_{( \frac{5}{2} )}}{r^{\frac{5}{2}}} + \frac{W_{(3 )}}{r^{3}} + \frac{W_{( \frac{7}{2} )}}{r^{\frac{7}{2}}} + \frac{ \textcolor{red}{ W_{(4)}}}{r^4}  +  \frac{\tilde{W}_{(4)}}{r^4} \ln r  + \cdots \nonumber \\
	{\cal U} &=& {\cal U}_{(0)} + \frac{ {\cal U}_{(1)}}{r} + \frac{ {\cal U}_{ ( \frac{3}{2} )}}{r^{\frac{3}{2}}} + \frac{ \textcolor{red}{{\cal U}_{(2)}}}{r^2} + \frac{\tilde{\cal U}_{(2)}}{r^2} \ln r +  \cdots \nonumber 
	\end{eqnarray}
Again, the integration functions are highlighted in colours, and all other coefficients in the expansions are expressed in terms of this data. The data $(\textcolor{ao(english)}{C_{i(0)}, \beta_{(0)}})$ is the analogue of non-normalisable boundary data in asymptotically locally AdS spacetimes, while the integration functions at subleading orders in the expansion $( \textcolor{blue}{C_{i ( \frac{3}{2} )}}, \textcolor{red}{W_{(4)}, {\cal U}_{(2)}})$ are analogous to normalisable boundary data. 

Explicit expressions for the first few terms in the expansions were derived in the previous section. These are summarised together with expressions for the subleading coefficients in the appendix \ref{appA}. Here we focus on the structure of these expansions and, in particular, how the logarithmic terms in these expansions arise.  

First of all, let us note that the integration functions in $C_i$ at order $r^{-3/2}$ are unaffected at this order by the new boundary conditions. Just as in previous works \cite{Tanabe:2009xb,Tanabe:2009va,Tanabe:2010rm}, we therefore expect that these integration functions are associated with gravitational waves: the defining data has the correct number of degrees of freedom to represent gravitational waves, and is unconstrained. Furthermore, as shown in appendix \ref{appA}, the $u$-evolution of the data $\textcolor{blue}{C_{i ( \frac{3}{2} )}}$ is left unspecified by the equations \eqref{Rsphere}, so that also $\partial_uC_{i( \frac{3}{2} )}$ have to be given as a coordinate on the phase space, in agreement with the asymptotically flat case. Note however that the new boundary conditions do affect the explicit forms for the expansion coefficients at subleading fractional orders. 

To definitely show that $C_{i (\frac{3}{2} )}$ corresponds to gravitational radiation, one would need to show that spacetimes with non-vanishing radiation lose mass. We postpone this question for future work, in which we will construct the conserved charges for asymptotically locally flat spacetimes. We should also note that, for $C_{i(\frac{3}{2})}$ to be interpreted as the degrees of freedom corresponding directly to gravitational waves, one should show rigorously that $\partial_uC_{i( \frac{3}{2} )}$ is gauge invariant, generalising the discussions of \cite{Satishchandran:2019pyc}. If $\partial_u C_{i(\frac{3}{2})}$ was not gauge invariant then one would need to construct a gauge invariant quantity that reduces to  $\partial_u C_{i(\frac{3}{2})}$ in the asymptotically flat case.

\bigskip

Consider the Einstein equations \eqref{beta2alpha} and \eqref{Rtrace2}; both these equations have associated integration functions, shown in red above. As discussed in the previous section, these equations can be viewed as inhomogeneous equations for $W$ and ${\cal U}$, respectively, which determine these functions from the functions that have already been determined. The asymptotic expansions of the functionals $({\cal F}, {\cal G})$ have the structure
\begin{eqnarray}
{\cal G} &=& \frac{{\cal G}_{(1)}}{r} + \frac{{\cal G}_{(2)}}{r^2} + \frac{{\cal G}_{(5/2)}}{r^{5/2}} + \frac{{\cal G}_{(3)}}{r^3} + 
\frac{{\cal G}_{(7/2)}}{r^{7/2}} + \frac{{\cal G}_{(4)}}{r^4} + \cdots \\
{\cal F} &=&  \frac{ {\cal F}_{(2)}}{r^2} +  \frac{ {\cal F}_{(3)}}{r^3}  +  \frac{ {\cal F}_{(7/2)}}{r^{7/2}}  + \frac{ {\cal F}_{(4)}}{r^4} + \cdots \nonumber
\end{eqnarray}	
Then integrating \eqref{beta2alpha} and \eqref{Rtrace2} we obtain
\begin{eqnarray}
\tilde{\cal U}_{(2)} &=& \frac{1}{3} {\cal F}_{(4)} e^{2 \beta_{(0)}} \\
\tilde{W}_4 &=& - \frac{1}{2} {\cal G}_{(4)} e^{2 \beta_{(0)} - C_{1 (0)}} \nonumber
\end{eqnarray}
Consistent integration of the Einstein equations thus requires either allowing for logarithmic terms in the asymptotic expansions or imposing constraints on the defining data such that ${\cal F}_{(4)} = {\cal G}_{(4)} = 0$. In the case of asymptotically flat spacetimes, the defining data is such that indeed these terms in the expansions of ${\cal F}$ and ${\cal G}$ vanish, so no logarithmic terms are induced. Note that while the terms $\tilde{\cal U}_{(2)}$ and $\tilde{W}_4$ are the leading logarithmic terms in the expansions, they clearly induce at subleading orders further logarithmic terms.

\bigskip

The occurrence of polyhomogenous asymptotic expansions is not new: they have arisen previously both for asymptotically flat spacetimes and for asymptotically locally anti-de Sitter spacetimes in odd dimensions. It is particularly useful to recall the case of asymptotically locally anti-de Sitter spacetimes in five dimensions, which solve the Einstein equations with cosmological constant and no matter. Then, working in Fefferman-Graham coordinates, the metric expansion in the vicinity of the conformal boundary $\rho \rightarrow 0$ is 
\begin{equation}
ds^2 = \frac{d \rho^2}{\rho^2} + \frac{1}{\rho^2} g_{ij} (x, \rho) dx^i dx^j
\end{equation} 	
where the four-dimensional metric $g_{ij}$ is expressed as 
\be
g_{ij} = \textcolor{ao(english)}{g_{(0) ij}} + \rho^2 g_{(2) ij} + \rho^4 \textcolor{red}{g_{(4) ij}} + \rho^4 \log \rho \; h_{(4) ij} + \cdots 
\ee	
Just as in the expansions given above, the data highlighted in colours completely determines the integration functions in solving the equations and hence the entire asymptotic expansion. All other terms in the expansion are expressed in terms of curvature tensors of $(\textcolor{ao(english)}{g_{(0)ij}}, \textcolor{red}{g_{(4)ij}})$. Explicit expressions for these coefficients may be found in \cite{Henningson:1998gx,deHaro2001} while the structure of asymptotically locally anti-de Sitter spacetimes and holographic renormalization are reviewed in \cite{Skenderis2002}. The interpretation of the defining data in the dual CFT is that the non-normalisable data $g_{(0) ij}$ is the background metric for the field theory, while the normalisable data $g_{(4) ij}$ determines the expectation value of the stress tensor in the CFT. An explicit expression for the stress tensor in terms of $g_{(4) ij}$ and covariant derivatives of $g_{(0) ij}$ can be found in \cite{deHaro2001}. The occurrence of logarithmic terms in the asymptotic expansion and in the regulated onshell action relates to Weyl anomalies in the field theory. 

The coefficient of the leading log term is  \cite{Henningson:1998gx,deHaro2001}
\begin{eqnarray}
h_{(4) ij} &=& \frac{1}{2} R_{(0)ikjl} R^{kl}_{(0)} - \frac{1}{12} \nabla_i \nabla_j R_{(0)} + \frac{1}{4} \nabla^2 R_{(0) ij} - \frac{1}{6} R_{(0)} R_{(0) ij} \label{h4} \\
&& - \frac{1}{24} \left ( \nabla^2 R_{(0)} - R_{(0)}^2 +3 R_{(0) kl} R^{kl}_{(0)} \right ) g_{(0) ij} \nonumber
\end{eqnarray}
where $\nabla$ is the covariant derivative associated with $g_{(0)}$ and $R_{(0)}$ denotes the curvature of $g_{(0)}$. Note that $h_{(4) ij}$ does not depend on the normalizable data $g_{(4) ij}$: conformal anomalies depend only on the background fields of the CFT, not on the specific state of the field theory and its energy momentum tensor. 

It is interesting to compare \eqref{h4} with our results for asymptotically locally flat spacetimes in five dimensions. In both cases, the coefficients of the leading log terms depend only on the non-normalizable data and its derivatives. If one restricts either asymptotically flat or to asymptotically AdS, the coefficients of the log terms vanish. 

The analysis of five-dimensional asymptotically locally anti-de Sitter spacetimes in Fefferman-Graham gauge makes manifest four-dimensional covariance, and accordingly this is the most commonly used gauge for asymptotic analysis. To make contact with our analysis of asymptotically locally flat spacetimes, one can instead use Bondi gauge for the asymptotic analysis in anti-de Sitter, see \cite{Poole:2018koa}
for the corresponding analysis in four dimensions. It would be interesting to carry out the asymptotic analysis in Bondi gauge for five dimensional asymptotically locally anti-de Sitter, and to explore the limit as the cosmological constant is taken to zero, in order to elucidate the structure found here.

	\bigskip

In the context of asymptotically flat spacetimes in four dimensions, polyhomogeneous expansions have been discussed in a number of earlier works, see \cite{Chrusciel-Friedrich,Kroon:1998tu,ValienteKroon:1998vn,ValienteKroon:1999cj,ValienteKroon:2000jy,ValienteKroon:2002gb}. In these contexts, however, the appearance of logarithmic terms is associated with non-smoothness of the boundary data; imposing suitable regularity conditions sets the logarithmic terms to zero. This fits with asymptotically locally anti-de Sitter spacetimes in four dimensions that satisfy Einstein's equations with negative cosmological constant: these also do not have logarithmic terms in their asymptotic expansions, since the Weyl anomaly of the stress tensor in a three-dimensional conformal field theory vanishes. More generally, it is well-known that the analysis of asymptotically flat spacetimes is very different in odd dimensions than in even dimensions; see recent discussions in \cite{Satishchandran:2019pyc}, which may also shed light on the polyhomogeneous terms in the expansions that arise here. 
	
\section{On asymptotic symmetries} \label{sec:six}

In this section we initiate the study of asymptotic symmetries of asymptotically locally flat spacetimes in five dimensions and comment on new features relative to previous analyses of the symmetry group of asymptotically flat spacetimes in five dimensions. 

Assuming that both $C_{i(0)}$ $h_{(0)AB}$ and $\beta_{(0)}$ in \eqref{ALFexpansion} are fixed as boundary data which partially specify the phase space, then the leading order terms in \eqref{leading_terms} are fixed, as can be checked from equations \eqref{W1} and \eqref{U0}. Together with the gauge preserving conditions
\begin{equation}\label{gauge}
\mathfrak{L}_{\xi}g_{rr}=\mathfrak{L}_{\xi}g_{rA}=0\,,\quad g^{AB}\mathfrak{L}_{\xi}g_{AB}=0\,,
\end{equation}
these asymptotics imply that the asymptotic Killing equations are
\be\label{killing}
\mathfrak{L}_{\xi}g_{AB}= \mathcal{O}(r) \,,\quad
\mathfrak{L}_{\xi}g_{uu}= \mathcal{O}(r^{-1}) \,, \quad \mathfrak{L}_{\xi}g_{ur}= \mathcal{O}(r^{-2})\,, \quad \mathfrak{L}_{\xi}g_{uA}= \mathcal{O}(r^0) \,.
\ee
In analysing solutions to these equations there is no benefit from restricting to $h_{AB}$ diagonal and $W^A=(W^{\q},0,0)$, and hence we consider here the general case without imposing additional symmetries. 

Given that the gauge preserving conditions are as in four dimensions, the asymptotic Killing vector fields also take the same form as in four dimensions \cite{Sachs1962}, namely 
\begin{equation}
\xi=\xi^u\partial_u+\xi^r\partial_r+\xi^A\partial_A
\end{equation}
with
\begin{equation}\label{killingVect}
\begin{cases}
\xi^u=f(u,x^A) \, ,\\
\xi^r= 	-\dfrac{r}{3}\left[D_A\xi^A-W^C\partial_Cf\right]\, ,\\
\xi^A=Y^A(u,x^B)-\partial_Bf\int_{r}^{\infty}dr'\,e^{2\beta}g^{AB}\,.
\end{cases}
\end{equation}
Here the functions $f(u,x^A)$ and $Y(u,x^A)$ are determined by \eqref{killing} and $D$ is the covariant derivative with respect to $h_{(0)AB}$. For the particular metric considered in this paper $W^C=W^\theta=:W$ and thus
\begin{equation}
D_A\xi^A=(\partial_\theta+(\cot\q-\tan\q))\xi^\theta+\partial_\phi\xi^\phi+\partial_\psi\xi^\psi\,.
\end{equation}

The right hand sides of \eqref{killing} have the same form as the corresponding four-dimensional equations. However, notice that while the first three conditions in \eqref{killing} are on the same footing of those of the four dimensional case, the latter is not. In five dimensions terms at order $r^0$ are subleading in the expansion of $g_{uA}$, while in the four-dimensional case such terms are leading and are determined by the shear tensor $C_{(1)AB}^{(4\text{dim})}$. 

Furthermore, in the present case the leading order terms of the expansions are different from the four dimensional asymptotically flat case, and hence there is a priori no guarantee that the symmetry algebra will have the same structure as in four dimensions i.e. the semi-direct sum of a non-Abelian part (either Lorentz or analogues of superrotations) and an Abelian part given by either translations or supertranslations. Indeed, in the case at hand the asymptotic Killing conditions determine $Y^A$ to be a  $u$-independent conformal Killing vector on the celestial sphere and the function $f(u,x^B)$ to have the general form
\begin{equation}\label{f}
f(u,x^B)=\frac{1}{3}F(u,x^B)+\tilde{\alpha}(u,x^B)
\end{equation}
where
\begin{align}\label{Falpha}
F(u,x^B)=e^{2\b_{(0)}}\int_{u} du&  e^{-2\b_{(0)}}\left(D_AY^A(x^B)-6\b_{(0),A}(u,x^B)Y^A(x^B)\right)\nonumber\\&
\tilde{\a}(u,x^B)=e^{2\b_{(0)}(u,x^B)}\alpha(x^B)
\end{align}
supplemented by other equations for the derivatives of $f$. We postpone the complete analysis of this system of equations and their consequences for the structure of the asymptotic symmetry group to the future.

To conclude this section we comment briefly on the differences between \eqref{Falpha} and the corresponding form of $f$ in four dimensions and in higher dimensions with asymptotically flat boundary conditions. In the latter cases $Y^A$ is a conformal Killing vector field, as in the present case, but the function $f(u,x^B)$ has the form in $d$ dimensions \cite{Tanabe2011}
\begin{equation}\label{bms}
f(u,x^B)=\frac{u}{d-2}D_AY^A(x^B)+\alpha(x^B)\, 
\end{equation}
where there is a split between a part which depends only on the angular coordinates and a part whose only dependence on $u$ is linear. This splitting is essential to get the finite form of the asymptotic symmetries as presented in \cite{Sachs1962} 
\begin{equation}
u\rightarrow \bar{u}=K^{-1}(x^B)[u+\alpha(x^B)],\qquad x^C\rightarrow\bar{x}^C=\Omega^{(C)}(x^D)
\end{equation} 
where $\Omega^{(C)}$ are conformal transformations on the angular coordinates and $K$ is the associated scale factor of the $S^{d-2}$ metric, from which the semi-direct product structure of the group is manifest. The Abelian part $\alpha$ gives supertranslations in four dimensions and only translations in higher spacetime dimensions according to the analysis of \cite{Tanabe2011}.


It is evident that if $\b_{(0)}$ is zero or constant
\eqref{Falpha} reduces to the standard form \eqref{bms}. As can be checked using the formulas in appendix \ref{appA}, such a choice of $\b_{(0)}$ does not collapse the asymptotic expansion of the metric to become asymptotically flat. Furthermore, one can prove that with $\mathcal{U}_{(0)}=1$ all the asymptotic Killing equations are satisfied without any constraint on $\mathcal{\a}$. This is in contrast to the asymptotically flat analysis in  \cite{Tanabe:2009va,Tanabe2011} where imposing $C_{i(0)}=0$ implies that $\mathfrak{L}_{\xi}g_{AB}=\mathcal{O}(\sqrt{r})$. This stronger condition provides one more equation which constrains $\alpha(x^A)$ to be a finite sum of scalar spherical harmonics with $l=0$ and $l=1$, forcing the asymptotic symmetry group to be the Poincar\'{e} group rather than the BMS group. Thus the asymptotic symmetry group with asymptotically locally flat boundary conditions reduces to that found in earlier literature when we restrict to asymptotically flat, but it is potentially a rich generalisation. 

We postpone for future work the detailed study of the asymptotic symmetry algebra, following the approach of \cite{Wald2000}. Given the general solution for the asymptotic Killing vectors \eqref{killingVect}, one needs to injects these vectors into the Wald-Zoupas symplectic form and thereby identify elements of the asymptotic symmetry algebra, factoring out degeneracies of the presymplectic product.

	\section{Conclusions and outlook} \label{sec:seven}
	
Motivated by the relation between cosmic strings and superrotations in four dimensions, we began this paper by exploring cosmic branes in higher dimensions. We argued that only $(d-3)$-branes in $d$ spacetime dimensions are flat in the vicinity of the brane, and therefore the natural generalization of cosmic strings/superrotations in four dimensions should involve $(d-3)$-branes. We then showed that, if one wishes to allow cosmic $(d-3)$-branes to penetrate the celestial sphere, one needs to relax the boundary conditions from asymptotically flat to asymptotically locally flat. 

 The proposed generalized boundary conditions are defined in \eqref{alfc} in terms of a non-trivial $(d-2)$ metric, describing a $(d-2)$-manifold that is topologically a $(d-2)$-sphere. These boundary conditions include cosmic branes, but the rather general form is primarily motivated by the analogy with asymptotically  locally anti-de Sitter spacetimes. The generalization of $d$-dimensional asymptotically anti-de Sitter spacetimes (for which the metric on the conformal boundary is ${\cal R}_{t} \times S^{d-2}$) to asymptotically locally anti-de Sitter spacetimes is obtained by allowing the metric on the conformal boundary is a generic smooth (non-degenerate) metric. As mentioned in the previous section, it would be interesting to explore the detailed relationship with anti-de Sitter, by expressing asymptotically locally anti-de Sitter spacetimes in Bondi gauge as in \cite{Poole:2018koa} and then taking the limit of zero cosmological constant. Related analysis in the special case of three spacetime dimensions was carried out in \cite{Bagchi:2012xr} and subsequent works.

We showed that the vacuum Einstein equations can be solved consistently with these boundary conditions near the null boundary. The resulting  
asymptotic expansions are polyhomogeneous, with the leading logarithmic terms in the expansions are expressed in terms of derivatives of the boundary metric on the celestial sphere. Again, this seems very much analogous to the structure of asymptotically locally anti-de Sitter spacetimes in five dimensions. 

There are many open questions that should be explored in future work. The original motivation for exploring cosmic branes was to relate them to the asymptotic symmetry group. One would thus like to derive the asymptotic symmetry group with the generalized boundary conditions, developing the analysis in section \ref{sec:six} and show whether the conserved charges associated with this group can be related to soft scattering theorems. With asymptotically flat boundary conditions, the integration constants at subleading order are proportional to the mass and angular momenta. With generalized boundary conditions, one should show that the mass and angular momenta are finite and derive explicit expressions for them using the methods of \cite{Wald:1993nt,Iyer:1994ys,Wald2000} together with the more recent work  \cite{Flanagan2018}. 

In the case of asymptotically flat spacetimes in five dimensions, the mass aspect is proportional to the coefficient of $g_{uu}$ at order $1/r^2$, i.e. the integration function ${\cal U}_{(2)}$. One would anticipate that the expression for the mass aspect $m_B$ in the general case will be much more complicated:
\be
m_{B} \sim {\cal U}_{(2)} + {\cal M} (\beta_{(0)}, C_{i(0)})
\ee
where ${\cal M}$ is a functional of the defining boundary data and its derivatives, which vanishes in the case of an asymptotically flat spacetime. An analogous structure was found for the mass aspect in asymptotically locally AdS spacetimes, see \cite{Poole:2018koa}. It would also be interesting to derive the explicit expression for the evolution in null time of the mass aspect.

 In this paper the asymptotic analysis was carried out for axisymmetric spacetimes with inversion symmetry and it would be interesting to extend this analysis to spacetimes without such symmetries. We would not expect the main conclusions of this work to change, i.e. the polyhomogeneous structure of the asymptotic expansion would persist, since the nested structure and integration of the Bondi gauge equations presented in section \ref{sec:four} does not rely on the symmetry assumptions. Relaxation of the symmetry assumptions gives additional equations for the other components of $W^A$ and $h_{AB}$, but these have a similar form to those given in section \ref{sec:four}, as for 5d asymptotically flat spacetimes discussed in  \cite{Tanabe:2009va,Tanabe:2010rm,Tanabe2011}.

We have shown by iterative integration that the vacuum Einstein equations can be solved consistently with the generalized boundary 
conditions. It would be interesting to proof the existence and uniqueness of such solutions rigorously, and to derive the convergence properties of the polyhomogeneous series. Note however that rigorous proofs and derivations may be challenging. In the case of asymptotically locally anti-de Sitter spacetimes, rigorous proofs of existence and uniqueness in Euclidean signature were given in the original mathematics literature \cite{FG1,GL}, but many outstanding issues still remain in Lorentzian signature. In the case of zero cosmological constant, the analysis is inherently Lorentzian and thus likely to be subtle. However, it was shown in \cite{Korovin:2017xqu} that the asymptotic analysis is simplified in first order formalism and this approach may be useful for rigorous proofs. 

BMS and extended BMS symmetry groups are relevant not just as asymptotic symmetry groups, but also as near-horizon symmetry groups for black holes.  In \cite{Shi2016} the BMS-like near-horizon symmetries were analysed in arbitrary dimensions. It would be interesting to explore the relations between these symmetry groups and the symmetry groups of asymptotically locally flat spacetimes.

As well as soft scattering theorems, another implication of extended BMS symmetries is memory effects. In particular, superrotations are associated with spin memory effects, see \cite{Flanagan2015,Flanagan2016,Nichols2017}. \cite{Nichols2017} studies the spin memory effect for compact binaries and emphasises that the spin memory effect is on a qualitatively different footing to the standard displacement memory effect (associated with supertranslations) because superrotated spacetimes are not strictly asymptotically flat even in four dimensions. While memory effects would manifestly not be observable in dimensions $d > 4$, it could nonetheless be elucidating to explore how memory effects manifest in higher dimensions. 

	\section*{Acknowledgements}
	
	This work is funded by EPSRC and by the STFC grant ST/P000711/1. This project has received funding and support from the European Union's Horizon 2020 research and innovation programme under the Marie Sklodowska-Curie grant agreement No 690575. MMT would like to thank CERN, the Kavli Institute for the Physics and Mathematics of the Universe and the Banff International Research Station for hospitality during the completion of this work. We thank Y.~Korovin, J.~Podolsky and K.~Skenderis for useful comments and discussions.  
	
	\appendix

	\section{Solutions of the main equations and supplementary equations} \label{appA}
	In this appendix we collect the solutions of the main equations as well as the supplementary equations. In writing the appendix a logistic problem concerning the typesetting of the equations arose: whether to write all equations in terms of the initial and free data or implicitly in terms of the previously determined data. We have used one form or the other according to space constraints; shorter equations are usually written in the fully expanded form while the longest ones are not.
	\subsection{$\beta$ coefficients} 
	\be
	\b_{(2)}=-\frac{1}{24}(C_{1(1)}^2-C_{1(1)}C_{2(1)}-C_{2(1)})
	\ee
	\be
	\b_{(5/2)}=-\frac{1}{20}(2 C_{1(1)} C_{1(3/2)}+C_{2(1)} C_{1(3/2)}+C_{1(1)} C_{2(3/2)}+2 C_{2(1)}C_{2(3/2)})
	\ee
	\begin{eqnarray}
	\b_{(3)}&=& -\frac{1}{16}\left( C_{1\left(\frac{3}{2}\right)}^2+
	C_{2\left(\frac{3}{2}\right)} C_{1\left(\frac{3}{2}\right)}+
	C_{2\left(\frac{3}{2}\right)}^2\right)\nonumber\\
	&&-\frac{1}{9}\left(C_{1(1)} C_{1(2)}+C_{2(1)} C_{2(2)}\right)\nonumber\\&&+\frac{1}{18}\left( C_{1(2)} C_{2(1)}+ C_{1(1)}
	C_{2(2)}\right)
	\end{eqnarray}
	
	
	

	\subsection{$W$ coefficients}
	\be\label{W1}
	W_{(1)} = 2e^{2 \beta_{(0)} - C_{1(0)}} \partial_{\theta} \beta_{(0)}
	\ee	
	\begin{align}
8e^{-2\b_{(0)}+C_{1(0)}}W_{(2)}&=2C_{1(0),\q} C_{1(1)} +
	C_{2(0)\q} C_{1(1)}+
	C_{1(0),\q} C_{2(1)}\nonumber\\& +2
	C_{2(0),\q} C_{2(1)}-2
	C_{1(1),\q} +2 \csc \q \sec\q C_{2(1)}  \nonumber\\&
+2 (2\tan \q-\cot\q)	C_{1(1)}-4 C_{1(1)} W_{(1)}
	\end{align}
	\begin{align}
	5 e^{-2 \b_{(0)}+C_{1(0)}} W_{(5/2)}&=2C_{1\left(\frac{3}{2}\right)} \left(- e^{-2 \b_{(0)}+C_{1(0)}}W_{(1)}+ C_{1(0),\q}C_{2(0),\q}\right)\nonumber\\&+4C_{1\left(\frac{3}{2}\right)}\left(2 \tan\q
	-\cot\q\right)-2 C_{1\left(\frac{3}{2}\right),\q} \nonumber\\& +C_{2\left(\frac{3}{2}\right)} \left( 
	\left(C_{1(0),\q}+2 C_{2(0),\q}\right)+2 \csc\q\sec
	\q\right)
	\end{align}
\begin{align}
3e^{-2\b_{(0)}+C_{1(0)}}W_{(3)}&=10 \b_{(2)}^{(0,1)}-2 C_{1(2),\q}+C_{1(1)}
C_{1(1),\q}\nonumber\\&+\frac{1}{2} C_{1(1)} C_{2(1),\q}+\frac{1}{2} C_{2(1)}
C_{1(1),\q}+ C_{2(1)} C_{2(1),\q}\nonumber\\&+2 C_{1(2)} C_{1(0),\q}+ C_{2(2)} C_{1(0),\q}+
C_{1(2)} C_{2(0),\q}\nonumber\\&+2 C_{2(2)} C_{2(0),\q}-2 C_{1(2)} (\cot\q
-2 \tan\q)+2 C_{2(2)} \csc\q \sec\q\nonumber\\&+ e^{C_{1(0)}-2
	\b_{(0)}}\left(2 \b_{(2)} W_{(1)}-\frac{1}{2}
(C_{1(1)})^2 W_{(1)}-2 C_{1(1)} W_{(2)}- C_{1(2)} W_{(1)}\right)
\end{align}
\begin{align}
7e^{-2\b+C_{(1)}} W_{(7/2)}&=44 \b_{\left(\frac{5}{2}\right),\q}-10
C_{1\left(\frac{5}{2}\right),\q}+4 C_{1(1)}
C_{1\left(\frac{3}{2}\right),\q}+2 C_{1(1)}
C_{2\left(\frac{3}{2}\right),\q}+6 C_{1\left(\frac{3}{2}\right)}
C_{1(1),\q}\nonumber\\&+3 C_{1\left(\frac{3}{2}\right)} C_{2(1),\q}+2 C_{2(1)}
C_{1\left(\frac{3}{2}\right),\q}+4 C_{2(1)}
C_{2\left(\frac{3}{2}\right),\q}+3 C_{2\left(\frac{3}{2}\right)}
C_{1(1),\q}\nonumber\\&+6 C_{2\left(\frac{3}{2}\right)} C_{2(1),\q}+10 C_{1\left(\frac{5}{2}\right)} C_{1(0),\q}+5 C_{2\left(\frac{5}{2}\right)} C_{1(0),\q}+5
C_{1\left(\frac{5}{2}\right)} C_{2(0),\q}\nonumber\\&+10 C_{2\left(\frac{5}{2}\right)}
C_{2(0),\q}-10 C_{1\left(\frac{5}{2}\right)} (\cot\q-2 \tan \q)+10 C_{2\left(\frac{5}{2}\right)} \csc\q\sec\q\nonumber\\&+e^{C_1(0)-2 \b(0)}\left(4
\b_{\left(\frac{5}{2}\right)} W_{(1)}-2 C_{1(1)} C_{1\left(\frac{3}{2}\right)}
W_{(1)}-2 C_{1\left(\frac{5}{2}\right)} W_{(1)}\right)\nonumber\\&-e^{C_1(0)-2 \b(0)}\left(5 C_{1(1)}
W_{\left(\frac{5}{2}\right)}+4 C_{1\left(\frac{3}{2}\right)} W_{(2)}\right)
\end{align}

At order $r^{-4}$ the equation determine the coefficient $\tilde{W}_{(4)}$ of the log term 
\begin{align}
16e^{-2\b+C_{1(0)}}\tilde{W}_{(4)}&=48 \b_{(3),\q}+12 C_{1(0),\q} C_{1(3)}+6 C_{2(0),\q}
C_{1(3)}+6 C_{1(0),\q} C_{2(3)}\nonumber\\&+12 C_{2(0),\q} C_{2(3)}-12 C_{1(3),\q}+4 C_{1(2),\q} C_{1(1)}+2
C_{2(2),\q} C_{1(1)}\nonumber\\&+6
C_{1\left(\frac{3}{2}\right),\q} C_{1\left(\frac{3}{2}\right)}+3
C_{2\left(\frac{3}{2}\right),\q} C_{1\left(\frac{3}{2}\right)}+8
C_{1(1),\q} C_{1(2)}+4 C_{2(1),\q} C_{1(2)}\nonumber\\&+2 C_{1(2),\q} C_{2(1)}+4 C_{2(2),\q}
C_{2(1)}+3 C_{1\left(\frac{3}{2}\right),\q}
C_{2\left(\frac{3}{2}\right)}+6 C_{2\left(\frac{3}{2}\right),\q}
C_{2\left(\frac{3}{2}\right)}\nonumber\\&+4 C_{1(1),\q} C_{{2(2)}}+8
C_{2(1),\q} C_{{2(2)}}-12 (2 \tan\q+\cot\q)
C_{1(3)}\nonumber\\&+12 \csc\q \sec\q C_{2(3)}
\end{align}
Here, substituting for $\b_{(3)}$ results in
	\begin{eqnarray}
	24e^{2\b-C_{1(0)}}\tilde{W}_{(4)}&=&-18 C_{1(3),\q}-2 C_{1(1)} C_{1(2),\q}-C_{1(1)}
	C_{2(2),\q}+4 C_{1(2)} C_{1(1),\q}\nonumber\\&&+2 C_{1(2)}
	C_{2(1),\q}-C_{2(1)} C_{1(2),\q}-2 C_{2(1)} C_{2(2),\q}+2
	C_{2(2)} C_{1(1),\q}\nonumber\\&&+4 C_{2(2)} C_{2(1),\q}+18 C_{1(3)} C_{1(0),\q}+9 C_{2(3) }C_{1(0),\q}+9 C_{1(3)} C_{2(0),\q}\nonumber\\&&+18 C_{2(3)}
	C_{2(0),\q}+36 C_{1(3)} \tan\q+18 C_{2(3)} \csc\q\sec\q-18 C_{1(3)}\qquad	
	\end{eqnarray}
	The subleading terms $W_{(k)}$  with $k>4$ can all be determined in terms of the previous ones as was the case up to $W_{(7/2)}$.
	
	\subsection{$\mathcal{U}$ coefficients}
	to be consistent with notation  are better written in a non-expanded form, but this implies a rweriting of all equations
	\begin{align}\label{U0}
	12 e^{2 \b_{(0)}+C_{1(0)}}\mathcal{U}_{(0)}&=4\b_{(0),\q}e^{4 \b_{(0)}}\left( C_{1(0),\q}- (\cot\q-\tan\q)-\b_{(0),\q}\right)\nonumber\\&+e^{4 \b_{(0)}}\left( \b_{(0),\q\q}-2(C_{1(0),\q})^2-(C_{2(0),\q})^2-C_{1(0),\q} C_{2(0),\q}+2 C_{1(0),\q\q}\right)\nonumber\\&+e^{4 \b_{(0)}}\left(5 (\cot\q -\tan\q ) C_{1(0),\q}-\csc\q  \sec \q \left(C_{1(0),\q}+2
	C_{2(0),\q}\right)+12\right)\nonumber\\&+10 W_{(1)}(\cot\q-\tan\q) e^{C_{1(0)}+2 \b_{(0)}}-e^{2 C_{1(0)}}W_{(1)}^2 
	\end{align}
	\begin{align}
	6 e^{2\b_{(0)}+C_{1(0)}}\mathcal{U}_{(1)}&= -e^{4 \beta_{(0)}}(4 C_{1(1),\q} C_{1(0),\q}+ C_{2(1),\q}
	C_{1(0),\q}+4C_{1(1)} \beta_{(0),\q} C_{1(0),\q}+)\nonumber\\&
 -e^{4 \beta_{(0)}}	C_{1(1),\q} C_{2(0),\q}-2 e^{4 \beta_{(0)}} C_{2(1),\q} C_{2(0),\q}+8
	W_{(2),\q} e^{C_{1(0)}+2 \beta_{(0)}}\nonumber\\&+ e^{4 \beta_{(0)}} (10C_{1(1),\q} +8\beta_{(0),\q}C_{1(1)} )   \cot 2\q-e^{4 \beta_{(0)}}
	C_{1(1),\q} \csc\q  \sec\q \nonumber\\&-2 e^{4 \beta_{(0)}} C_{2(1),\q} \csc\q\sec\q +4 e^{4 \beta_{(0)}} \beta_{(0),\q} C_{1(1),\q}+2 e^{4 \beta(0)} C_{1(1),\q\q}\nonumber\\&+4 e^{4 \beta_{(0)}} C_{1(1)}(\beta_{(0),\q})^2+4 e^{4 \beta
		(0)} C_{1(1)} \beta_{(0),\q\q}+2 e^{4 \beta_{(0)}} C_{1(1)}\nonumber\\& +(C_{1(0),\q})^2+e^{4
		\beta_{(0)}} C_{1(1)}(C_{2(0),\q})^2+e^{4 \beta_{(0)}} C_{1(1)} C_{1(0),\q}
	C_{2(0),\q}\nonumber\\&-10 e^{4 \beta_{(0)}} C_{1(1)} \cot 2 \theta C_{1(0),\q}+e^{4 \beta_{(0)}}
	C_{1(1)} \csc\q  \sec\q C_{1(0),\q}\nonumber\\&+2 e^{4 \beta_{(0)}} C_{1(1)} \csc \q
	\sec\q  C_{2(0),\q}-2 e^{4 \beta_{(0)}} C_{1(1)} C_{1(0),\q\q}\nonumber\\&-4 W_{(2)}
	e^{C_{1(0)}} \left(W_{(1)} e^{C_{1(0)}}-4 e^{2 \beta_{(0)}} \cot2 \theta\right)\nonumber\\&-C_{1(1)} W_{(1)}^2e^{2 C_{1(0)}}-12 e^{4 \beta_{(0)}} C_{1(1)}
	\end{align}
	\begin{align}
	3e^{-2 \b_{(0)}+C_{1(0)}}\mathcal{U}_{(3/2)}&= (-4 C_{1(0),\q} C_{1(\frac{3}{2}),\q}-C_{2(0),\q} C_{1(\frac{3}{2}),\q}-C_{1(0),\q}
	C_{2(\frac{3}{2}),\q})\nonumber\\&+(-2 C_{2(0),\q} C_{2(\frac{3}{2}),\q}+2 (C_{1(0),\q})^2
	C_{1(\frac{3}{2})}+(C_{2(0),\q})^2C_{1(\frac{3}{2})})\nonumber\nonumber\\&+ (C_{1(0),\q} C_{2(0),\q}
	C_{1(\frac{3}{2})}-4 C_{1(0),\q}\b_{(0),\q} C_{1(\frac{3}{2})})\nonumber\\&+\csc 2 \theta(2 C_{1(0),\q}
	C_{1(\frac{3}{2})}+4 C_{2(0),\q} C_{1(\frac{3}{2})}-2 C_{1(\frac{3}{2}),\q}-4
	C_{2(\frac{3}{2}),\q})\nonumber\\&(-2 C_{1(0),\q\q} C_{1(\frac{3}{2})}+4 \b_{(0),\q}
	C_{1(\frac{3}{2}),\q}+2 C_{1(\frac{3}{2}),\q\q}-12 e^{4 \b_{(0)}} C_{1(\frac{3}{2})})\nonumber\\&+4(
	(\b_{(0),\q})^2 C_{1(\frac{3}{2})}+4 \b_{(0),\q\q} C_{1(\frac{3}{2})})\nonumber\\&+\cot 2 \theta(-10 C_{1(0),\q} C_{1(\frac{3}{2})}+10  C_{1(\frac{3}{2}),\q}+8 \b_{(0),\q} C_{1(\frac{3}{2})})\nonumber\\&+e^{-2 \b_{(0)}+C_{1(0)}}(14
	W_{(\frac{5}{2})}+7 W_{(\frac{5}{2}),\q})\nonumber\\&-e^{-2\b_{(0)}+2 C_{1(0)}}(
	C_{1(\frac{3}{2})} (W_{(1)})^2-5 W_{(\frac{5}{2})} W_{(1)})
	\end{align}
\begingroup\allowdisplaybreaks
\begin{align}
12e^{2\b_{(0)}+C_{1(0)}}\tilde{\mathcal{U}}_{(2)}&=e^{2 \b_{(0)}+C_{1(0)}} (48 \b_{(2)} U_{(0)}-\cot 2\q(80 \b_{(2)} W_{(1)}+24 W_{(3)}))\nonumber\\&+e^{2 \b_{(0)}+C_{1(0)}}(-40 \b_{(2)} W_{(1),\q}+12 W_{(3),\q})\nonumber\\&+e^{2 C_{1(0)}} (-C_{1(1)}^2 W_{(1)}^2+8 \b_{(2)} W_{(1)}^2-2 C_{1(2)}
	W_{(1)}^2)\nonumber\\&+e^{2 C_{1(0)}} (-12 W_{(3)} W_{(1)}-8 W_{(2)} C_{1(1)} W_{(1)}-8 W_{(2)}^2)\nonumber\\&+e^{4 \b_{(0)}} (-4 C_{1(1)}^2 (\b_{(0),\q})^2+8 C_{1(2)} (\b_{(0),\q})^2-16 \b_{(2),\q} \b_{(0),\q})\nonumber\\&+e^{4 \b_{(0)}} (8 C_{1(2),\q} \b_{(0),\q}-8
	C_{1(1),\q} C_{1(1)} \b_{(0),\q}+4 C_{1(1)}^2 C_{1(0),\q} \b_{(0),\q})\nonumber\\&+e^{4 \b_{(0)}} (-8 C_{1(2)} C_{1(0),\q} \b_{(0),\q}-4 (C_{1(1),\q})^2-2 (C_{2(1),\q}){}^2-4 \b_{(0),\q\q} C_{1(1)}^2)\nonumber\\&+e^{4 \b_{(0)}} (12
	C_{1(1)}^2-2 C_{1(1)}^2 (C_{1(0),\q})^2+4 C_{1(2)} (C_{1(0),\q})^2-C_{1(1)}^2 (C_{2(0),\q})^2)\nonumber\\&
	+e^{4 \b_{(0)}} (2 C_{1(2)} (C_{2(0),\q})^2+ \csc 2\q(-4 C_{1(2),\q}-8 C_{2(2),\q}+4 C_{1(1),\q} C_{1(1)}))\nonumber\\&+e^{4 \b_{(0)}} (-2 C_{1(1),\q} C_{2(1),\q}-8 \b_{(2),\q\q}+4 C_{1(2)}{,\q\q}+8 \csc 2\q C_{2(1),\q} C_{1(1)})\nonumber\\&+e^{4 \b_{(0)}}(-4
	C_{1(1)}{,\q\q} C_{1(1)}+8 \b_{(0),\q\q} C_{1(2)}-24 C_{1(2)}-2 \csc 2\q C_{1(1)}^2 C_{1(0),\q})\nonumber\\&+e^{4 \b_{(0)}}(8 \b_{(2),\q} C_{1(0),\q}-8 C_{1(2),\q} C_{1(0),\q}-2 C_{2(2),\q} C_{1(0),\q}+8
	C_{1(1),\q} C_{1(1)} C_{1(0),\q})\nonumber\\&+e^{4 \b_{(0)}}(2 C_{2(1),\q} C_{1(1)} C_{1(0),\q}+4 \csc 2\q C_{1(2)} C_{1(0),\q})\nonumber\\&+e^{4 \b_{(0)}}\cot 2\q (-8 \b_{(0),\q} C_{1(1)}^2+10 C_{1(0),\q} C_{1(1)}^2-20
	C_{1(1),\q} C_{1(1)})\nonumber\\&+e^{4 \b_{(0)}}(-16 \b_{(2),\q}+20 C_{1(2),\q}+16 \b_{(0),\q} C_{1(2)}-20 C_{1(2)} C_{1(0),\q})\nonumber\\&+e^{4 \b_{(0)}}(-4 \csc 2\q C_{1(1)}^2 C_{2(0),\q}-2 C_{1(2),\q} C_{2(0),\q}-4 C_{2(2),\q}
	C_{2(0),\q})\nonumber\\&e^{4 \b_{(0)}}(+2 C_{1(1),\q} C_{1(1)} C_{2(0),\q}+4 C_{2(1),\q} C_{1(1)} C_{2(0),\q}+8 \csc 2\q C_{1(2)} C_{2(0),\q})\nonumber\\&e^{4 \b_{(0)}}(-C_{1(1)}^2 C_{1(0),\q} C_{2(0),\q}+2 C_{1(2)} C_{1(0),\q}
	C_{2(0),\q}+2 C_{1(1)}^2 C_{1(0),\q\q}-4 C_{1(2)} C_{1(0),\q\q})
\end{align}
\endgroup
The next equations determine $\mathcal{U}_{(i)}$ with $i\ge 7/2$ and as said in the main text $\mathcal{U}_{(2)}$ remains free.

\subsection{$C_{i,u}$ coefficients}
We collect here the derivatives with respect to $u$ of $C_i(n)$ determined by the equations \eqref{Rsphere}.
		
At order $r^0$:
\begin{align}
C_{1(1),u}={\cal H}_1&=\frac{1}{3} e^{2 \b_{(0)}-C_{1(0)}} \left((C_{1(0),\q})^2+2 C_{2(0),\q}
C_{1(0),\q}+2(C_{2(0),\q})^2-C_{1(0),\q\q}\right)\nonumber\\&
+\frac{2}{3} e^{2 \b_{(0)}-C_{1(0)}}\left(
C_{1(0),\q}  \b_{(0),\q}-8
( \b_{(0),\q})^2-4 \b_{(0),\q\q}\right)\nonumber\\&+\frac{1}{3} e^{2 \b_{(0)}-C_{1(0)}}(\tan \q-\cot \q) \left(C_{1(0),\q}-4 \b_{(0),\q}\right)\nonumber\\&
+\frac{2}{3} e^{2 \b_{(0)}-C_{1(0)}}\csc\q \sec\q \left( C_{1(0),\q}+2
C_{2(0),\q}\right)
\end{align}
and
\begin{align}
C_{2(1),u}={\cal H}_2&= \frac{1}{3} e^{2 \b_{(0)}-C_{1(0)}} \left(-4\b_{(0),\q} C_{1(0),\q}-6 \b_{(0),\q}
C_{2(0),\q}\right)\nonumber \\& + \frac{1}{3} e^{2 \b_{(0)}-C_{1(0)}}\left( +8
(\b_{(0),\q})^2+4 \b_{(0),\q\q}-2
C_{1(0),\q}^2\right)\nonumber\\&-\frac{2}{3} e^{2 \b_{(0)}-C_{1(0)}}(\tan\q+\cot\q)C_{1(0),\q}\nonumber\\&+\frac{1}{3}e^{2 \b_{(0)}-C_{1(0)}}(\cot\q-5 \tan\q) C_{2(0),\q} \nonumber\\&-\frac{4}{3}e^{2 \b_{(0)}-C_{1(0)}}( \tan\q+ 2\cot\q)
\b_{(0),\q}\nonumber\\&
-\frac{1}{3} e^{2 \b_{(0)}-C_{1(0)}}\left((C_{2(0),\q})^2+4 C_{1(0),\q} C_{2(0),\q}\right)\nonumber\\&
+\frac{1}{3} e^{2 \b_{(0)}-C_{1(0)}}\left(2 C_{1(0),\q\q}3
C_{2(0),\q\q}\right)
\end{align}
		
There are no equations at order $r^{-1/2}$. In particular no equations constrain $C_{i(\frac{3}{2}),u}$ and the next derivative to be determined is $C_{1(2),u}$ from the order $r^{-1}$:
\begin{align}
C_{1(2),u}&=-e^{2 \b_{(0)}-C_{1(0)}}\left(2 C_{1(1),\q}  C_{1(0),\q}+C_{2(1),\q}C_{1(0),\q}+C_{1(1),\q}C_{2(0),\q}+2 C_{2(1),\q}C_{2(0),\q}\right)\nonumber\\&
+e^{2\b_{(0)}-C_{1(0)}}\left(2 \b_{(0),\q} C_{1(1),\q}+C_{1(1),\q\q}+2
C_{1(1),\q} (\cot\q-\tan\q)-2 C_{1(1)} W_{(1)}^2\right)\nonumber\\&
-e^{2 \b_{(0)}-C_{1(0)}}\csc\q \sec\q\left(C_{1(1),\q}+2 C_{2(1),\q}\right)\nonumber\\& +5 C_{1(1)} W_{(1),\q}+C_{1(1)} C_{1(1),u}+4
W_{(2),\q}\nonumber\\&
+W_{(2)} \left(-4 W_{(1)} e^{C_{1(0)}-2 \b_{(0)}}+C_{1(0),\q}+2(\cot\q-\tan\q)\right)\nonumber\\&+2 C_{1(1)} W_{(1)} C_{1(0),\q}-3 C_{1(1)} \mathcal{U}_{(0)}+C_{1(1)} W_{(1)}
(\cot\q -\tan \q)-2 \mathcal{U}_{(1)}
\end{align}
\begin{align}
C_{2(2),u}&= e^{2 \b_{(0)}-C_{1(0)}}(C_{2(1),\q} C_{1(0),\q}+C_{1(1),\q}C_{2(0),\q}+2 C_{1(1)}
\b_{(0),\q}C_{2(0),\q}-2 C_{2(1)} \b_{(0),\q} C_{2(0),\q})\nonumber\\&+\tan\q \left(C_{2(1),\q} e^{2 \b_{(0)}-C_{1(0)}}-C_{1(1)} e^{2 \b_{(0)}-C_{1(0)}} C_{2(0),\q}+C_{2(1)} e^{2
	\b_{(0)}-C_{1(0)}} C_{2(0),\q}\right)\nonumber\\&+\cot\q e^{2 \b_{(0)}-C_{1(0)}} \left(4 C_{1(1)} \b_{(0),\q}-4 C_{2(1)}
\b_{(0),\q}+2 C_{1(1),\q}-C_{2(1),\q}-2 C_{1(1)}C_{1(0),\q}\right)\nonumber\\&+\cot\q e^{2 \b_{(0)}-C_{1(0)}} \left(+2 C_{2(1)} C_{1(0),\q}+C_{1(1)} C_{2(0),\q}-C_{2(1)}
C_{2(0),\q}\right)\nonumber\\&+e^{2 \b_{(0)}-C_{1(0)}}(
-2 \b_{(0),\q} C_{2(1),\q} -C_{2(1),\q\q})+C_{2(1)}
W_{(1),\q}+C_{2(1)} C_{2(1),u}\nonumber\\&+2 W_{(2),\q}+e^{2 \b_{(0)}-C_{1(0)}}(-C_{1(1)} C_{1(0),\q} C_{2(0),\q}+C_{2(1)} C_{1(0),\q} C_{2(0),\q}+C_{1(1)} C_{2(0),\q\q})\nonumber\\&+2 W_{(2),\q}+e^{2 \b_{(0)}-C_{1(0)}}(-C_{2(1)} C_{2(0),\q\q}-4 C_{1(1)}+4 C_{2(1)})\nonumber\\&+\left(-4 C_{2(1)} W_{(1)}-3 W_{(2)}\right) \tan\q+\left(2 C_{2(1)} W_{(1)}+3 W_{(2)}\right)
\cot\q\nonumber\\&
+W_{(2)} C_{2(0),\q}+2 C_{2(1)} W_{(1)} C_{2(0),\q}-3 C_{2(1)} \mathcal{U}_{(0)}\nonumber\\&+\left(3 C_{2(1)} W_{(1)}+W_{(2)}\right) \csc\q \sec\q-2 \mathcal{U}_{(1)}
\end{align}
The equations at subleading orders determine the $u$-derivatives of $C_{i(n/2)}$ with $n>3$. In general, the equation for $C_{i(n/2),u}$ with $n\in\mathbb{N}\smallsetminus 3$ comes from the order $r^{n/2-1}$.
\subsection{Supplementary equations}
In order for the Bondi procedure to be consistent, once the main equations are solved the supplementary equations turns out to be automatically solved except at the order in which the free integration functions $\mathcal{U}_{(2)}$ and $W_{(4)}$ enter, in which case they give their evolution equation. These are the following.

At order $r^{-3}$ in $R_{u\q}=0$
\begingroup\allowdisplaybreaks
	\begin{eqnarray}
	3(\mathcal{U}_{(2)}-4\log r \tilde{\mathcal{U}}_{(2)})_{,u}&=&24\log r \tilde{\mathcal{U}}_{(2)}\b_{(0),u}+ W_{(1)}+(C_{1(\frac{3}{2}),u}){}^2+
	(C_{2(\frac{3}{2}),u}){}^2+\mathcal{U}_{(0)} \mathcal{U}_{(1)}\nonumber\\&&+2 \mathcal{U}_{(0)} W_{(2)} \b_{(0),\q}
	+W_{(2)}\mathcal{U}_{(0),\q}-\mathcal{U}_{(1)} W_{(1),\q}-4 W_{(2)} \b_{(0),\q} W_{(1),\q}\nonumber\\&&+4 W_{(1),\q}
	W_{(2),\q}+2 W_{(2)} W_{(1),\q\q}+6 U_{(2)}
	\b_{(0),u}-4 W_{(3),\q} \b_{(0),u}\nonumber\\&&-2 \mathcal{U}_{(0)} \b_{(2),u}-4 W_{(1),\q}
	\b_{(2),u}+2 W_{(2),\q} C_{1(1),u}+2 W_{(1),\q} C_{1(2),u}\nonumber\\&&+2
	C_{1(1),u} C_{1(2),u}+C_{1(2),u}
	C_{2(1),u}+C_{1(\frac{3}{2}),u}
	C_{2(\frac{3}{2}),u}+C_{1(1),u} C_{2(2),u}\nonumber\\&&+2
	C_{2(1),u} C_{2(2),u}-4 W(3) \b_{(0),u\q}+2 W(3)^{(1,1)}+2 W_{(2)}
	C_{1(1),u\q}\nonumber\\&&+e^{C_{1(0)}-2 \b_{(0)}} (\tan\theta -\cot \theta )
	(C_{1(1)} W_{(1)}^3+2 W_{(2)} W_{(1)}^2)\nonumber\\&&+e^{2 C_{1(0)}-4 \b_{(0)}} (2
	C_{1(1)} W_{(1)}^4+6 W_{(2)} W_{(1)}^3)+2 W_{(2)} W_{(1),\q}
	C_{1(0),\q}\nonumber\\&&+e^{2 \b_{(0)}-C_{1(0)}} (\cot\theta
	-\tan \theta ) (2 \mathcal{U}_{(1)} \b_{(0),\q}-2 \mathcal{U}_{(0)} C_{1(1)}
	\b_{(0),\q})\nonumber\\&&
	+e^{2 \b_{(0)}-C_{1(0)}} (\cot\theta
	-\tan \theta )(\mathcal{U}_{(1),\q}-\mathcal{U}_{(0),\q} C_{1(1)})\nonumber\\&&
	+(\cot \theta -\tan \theta ) (-C_{1(1),\q}
	W_{(1)}^2+\mathcal{U}_{(1)} W_{(1)}-2 W_{(2),\q} W_{(1)}+4 \b_{(2),u} W_{(1)})\nonumber\\&&-2(\cot \theta -\tan \theta )( C_{1(2),u}
	W_{(1)}+ W_{(2)} C_{1(0),\q} W_{(1)}+ W_{(2)} W_{(1),\q})\nonumber\\&&+2(\cot \theta -\tan \theta )(2 W_{(3)} \b_{(0),u}-
	W_{(3),u}- W_{(2)} C_{1(1),u})\nonumber\\&&+e^{C_{1(0)}-2 \b_{(0)}} (2
	C_{1(1),\q} W_{(1)}^3-4 \b_{(0),\q} C_{1(1)} W_{(1)}^3-2 \b_{(0),u} C_{1(1)}{}^2
	W_{(1)}^2)\nonumber\\&&+e^{C_{1(0)}-2 \b_{(0)}} (-3 \mathcal{U}_{(1)} W_{(1)}^2-16 W_{(2)} \b_{(0),\q} W_{(1)}^2+4 W_{(2),\q} W_{(1)}^2)\nonumber\\&&+e^{C_{1(0)}-2 \b_{(0)}} (+8 \b_{(2)}
	\b_{(0),u} W_{(1)}^2-4 \b_{(2),u} W_{(1)}^2+3 C_{1(2),u}
	W_{(1)}^2)\nonumber\\&&+e^{C_{1(0)}-2 \b_{(0)}} (W_{(1),\q} C_{1(1)} W_{(1)}^2+C_{1(1),u} C_{1(1)} W_{(1)}^2-4 \b_{(0),u}
	C_{1(2)} W_{(1)}^2)\nonumber\\&&+e^{C_{1(0)}-2 \b_{(0)}} (3 W_{(2)} C_{1(0),\q} W_{(1)}^2+W_{(1),u} C_{1(1)}{}^2 W_{(1)}-2
	\mathcal{U}_{(0)} W_{(2)} W_{(1)})\nonumber\\&&+e^{C_{1(0)}-2 \b_{(0)}} (+4 W_{(2)} W_{(1),\q} W_{(1)}-16 W(3) \b_{(0),u} W_{(1)}-4 \b_{(2)}
	W_{(1),u} W_{(1)})\nonumber\\&&+e^{C_{1(0)}-2 \b_{(0)}} (6 W_{(3),u} W_{(1)}+4 W_{(2)} C_{1(1),u} W_{(1)}-12 W_{(2)}
	\b_{(0),u} C_{1(1)} W_{(1)})\nonumber\\&&+e^{C_{1(0)}-2 \b_{(0)}} (4 W_{(2),u} C_{1(1)} W_{(1)}+2 W_{(1),u} C_{1(2)}
	W_{(1)}-8 W_{(2)}^2 \b_{(0),u})\nonumber\\&&+e^{C_{1(0)}-2 \b_{(0)}} (2 W(3) W_{(1),u}+4 W_{(2)} W_{(2),u}+2 W_{(2)}
	W_{(1),u} C_{1(1)})\nonumber\\&&+e^{2 \b_{(0)}-C_{1(0)}} (-4 \mathcal{U}_{(1)}
	(\b_{(0),\q})^2+4 \mathcal{U}_{(0)} C_{1(1)} (\b_{(0),\q})^2)\nonumber\\&&+e^{2 \b_{(0)}-C_{1(0)}} (-2
	\mathcal{U}_{(1),\q} \b_{(0),\q}+2 \mathcal{U}_{(0)} C_{1(1),\q} \b_{(0),\q}+2 \mathcal{U}_{(0),\q}
	C_{1(1)} \b_{(0),\q})\nonumber\\&&+e^{2 \b_{(0)}-C_{1(0)}} (2 \mathcal{U}_{(1)} C_{1(0),\q} \b_{(0),\q}-2 \mathcal{U}_{(0)} C_{1(1)}
	C_{1(0),\q} \b_{(0),\q}+\mathcal{U}_{(0),\q} C_{1(1),\q})\nonumber\\&&+e^{2 \b_{(0)}-C_{1(0)}} (-2 \mathcal{U}_{(1)}
	\b_{(0),\q\q}-\mathcal{U}_{(1),\q\q}+2 \mathcal{U}_{(0)} \b_{(0),\q\q} C_{1(1)})\nonumber\\&&+e^{2 \b_{(0)}-C_{1(0)}} (\mathcal{U}_{(0),\q\q}
	C_{1(1)}+\mathcal{U}_{(1),\q} C_{1(0),\q}-\mathcal{U}_{(0),\q} C_{1(1)} C_{1(0),\q})\,,
	\end{eqnarray}
	and at order $r^{-7/2}$ in $R_{u\q}=0$
	\begin{align}
	24W_{\left(4\right),u}&=-2 \log r \left(-36 \tilde{\mathcal{U}}_{(2)} W_{(1)}-12 e^{2 \beta_{(0)}-C_{1(0)}} \tilde{\mathcal{U}}_{(2),\q}+48 \tilde{W}_{(4),u}-96 \tilde{W}_{(4)} \beta_{(0),u}\right)\nonumber\\&-3\cot\q \left( C_{1(1)}{}^2 W_{(1)}^2+4 \b_{(2)} W_{(1)}^2+2 C_{1(2)} W_{(1)}^2+4 e^{2 \beta_{(0)}-C_{1(0)}} \b_{(2),\q} W_{(1)}\right)\nonumber\\&-3\cot\q\left(2 W_{(2)} C_{1(1)} W_{(1)}+4 e^{2 \beta_{(0)}-C_{1(0)}} W_{(3)} \beta_{(0),\q}+3 e^{2 \beta_{(0)}-C_{1(0)}}
	C_{1(3),u}\right)\nonumber\\&+3 \tan\q \left( C_{1(1)}^2 W_{(1)}^2+4\b_{(2)} W_{(1)}^2+2 C_{1(2)} W_{(1)}^2+4 e^{2 \beta_{(0)}-C_{1(0)}} \b_{(2),\q} W_{(1)}\right)\nonumber\\&+3 \tan\q \left(2W_{(2)} C_{1(1)} W_{(1)}+4 e^{2 \beta
		(0)-C_{1(0)}} W_{(3)} \beta_{(0),\q}+3 e^{2 \beta_{(0)}-C_{1(0)}} C_{1(3),u}\right)\nonumber\\&-3 e^{C_{1(0)}-2 \beta_{(0)}} \left( 4C_{1(1)}^2 W_{(1)}^3-8 \b_{(2)} W_{(1)}^3+4 C_{1(2)} W_{(1)}^3+14 W_{(3)}
	W_{(1)}^2\right)\nonumber\\&-3 e^{C_{1(0)}-2 \beta_{(0)}} \left(20 W_{(2)} C_{1(1)} W_{(1)}^2+16 W_{(2)}^2 W_{(1)}\right)\nonumber\\&-
12	e^{2 \beta_{(0)}-C_{1(0)}} \left( W_{(3)} \left(\beta_{(0),\q}\right)^2+4 W_{(1)} \b_{(2),\q} \beta_{(0),\q}- W_{(2)} C_{1(1),\q} \beta_{(0),\q}\right)\nonumber\\&+
	6e^{2 \beta_{(0)}-C_{1(0)}} \left(2W_{(1)} C_{1(2),\q} \beta_{(0),\q}+2 W_{(3)}
	C_{1(0),\q} \beta_{(0),\q}- \tilde{\mathcal{U}}_{(2),\q}+4 \b_{(2)} \mathcal{U}_{(0),\q}\right)\nonumber\\&+
	3e^{2 \beta_{(0)}-C_{1(0)}} \left(2 \mathcal{U}_{(2),\q}-4W_{(3)} \beta_{(0),\q\q}-4 W_{(1)} \b_{(2),\q\q}+2 C_{1(2),\q} C_{1(1),u}\right)\nonumber\\&+
	3e^{2 \beta_{(0)}-C_{1(0)}} \left(C_{2(2),\q}
	C_{1(1),u}+2 C_{1\left(\frac{3}{2}\right),\q} C_{1\left(\frac{3}{2}\right),u}+ C_{2\left(\frac{3}{2}\right),\q} C_{1\left(\frac{3}{2}\right),u}\right)\nonumber\\&+
	3e^{2 \beta_{(0)}-C_{1(0)}} \left(2 C_{1(1),\q} C_{1(2),u}+
	C_{2(1),\q} C_{1(2),u}+C_{1(2),\q} C_{2(1),u}\right)\nonumber\\&+
	3e^{2 \beta_{(0)}-C_{1(0)}} \left(2 C_{2(2),\q} C_{2(1),u}+C_{1\left(\frac{3}{2}\right),\q} C_{2\left(\frac{3}{2}\right),u}+2
	C_{2\left(\frac{3}{2}\right),\q} C_{2\left(\frac{3}{2}\right),u}\right)\nonumber\\&+
	3e^{2 \beta_{(0)}-C_{1(0)}} \left( C_{1(1),\q} C_{2(2),u}+2 C_{2(1),\q} C_{2(2),u}+ \csc\q  \sec\q \left(C_{1(3),u}+2
	C_{2(3),u}\right)\right)\nonumber\\&+
	6e^{2 \beta_{(0)}-C_{1(0)}} \left(2 \b_{(3),\q\q}-6 C_{1(3),\q\q}- \mathcal{U}_{(1),\q} C_{1(1)}-2 \mathcal{U}_{(0),\q} C_{1(2)}+2 W_{(1)} \b_{(2),\q} C_{1(0),\q}\right)\nonumber\\&+
	3e^{2 \beta_{(0)}-C_{1(0)}} \left(2 C_{1(3),u} C_{1(0),\q}+ C_{2(3),u}
	C_{1(0),\q}+ C_{1(3),u} C_{2(0),\q}+2 C_{2(3),u} C_{2(0),\q}\right)\nonumber\\&-W_{(1),u} C_{1(1)}^3+2 W_{(1)} \beta_{(0),u} C_{1(1)}^3-3 U_{(0)} W_{(1)} C_{1(1)}^2+3 W_{(1)} W_{(1),\q} C_{1(1)}^2\nonumber\\&+6 W_{(1)}^2 \beta_{(0),\q}
	C_{1(1)}^2-6 W_{(2),u} C_{1(1)}^2+12 W_{(2)} \beta_{(0),u} C_{1(1)}^2+3 W_{(1)}^2 C_{1(0),\q} C_{1(1)}^2\nonumber\\&-6 \mathcal{U}_{(0)} W_{(2)} C_{1(1)}-6 W_{(1)} W_{(2),\q} C_{1(1)}+36 W_{(1)} W_{(2)} \beta_{(0),\q} C_{1(1)}-6 W_{(1)}^2 C_{1(1),\q}
	C_{1(1)}\nonumber\\&+12 \b_{(2)} W_{(1),u} C_{1(1)}-18 W_{(3),u} C_{1(1)}+36 W_{(3)} \beta_{(0),u} C_{1(1)}-24 W_{(1)} \b_{(2)} \beta_{(0),u} C_{1(1)}\nonumber\\&+12 W_{(1)} \b_{(2),u} C_{1(1)}-6 W_{(2)} C_{1(1),u} C_{1(1)}-12 W_{(1)}
	C_{1(2),u} C_{1(1)}-6 W_{(1),u} C_{1(2)} C_{1(1)}\nonumber\\&+12 W_{(1)} \beta_{(0),u} C_{1(2)} C_{1(1)}-3 W_{(1),u} C_{1\left(\frac{3}{2}\right)}{}^2+6 W_{(1)} \beta_{(0),u} C_{1\left(\frac{3}{2}\right)}{}^2-12 \tilde{\mathcal{U}}_{(2)} W_{(1)}\nonumber\\&+18
	\mathcal{U}_{(2)} W_{(1)}+12 \mathcal{U}_{(1)} W_{(2)}-6 U_{(0)} W_{(3)}+36 U_{(0)} W_{(1)} \b_{(2)}-6 W_{(3)} W_{(1),\q}\nonumber\\&-36 W_{(1)} \b_{(2)} W_{(1),\q}-12 W_{(2)} W_{(2),\q}-18 W_{(1)} W_{(3),\q}+24 W_{(2)}^2 \beta_{(0),\q}\nonumber\\&+48 W_{(1)} W_{(3)} \beta_{(0),\q}-24 W_{(1)}^2
	\b_{(2)} \beta_{(0),\q}+12 W_{(1)}^2 \b_{(2),\q}-24 W_{(1)} W_{(2)} C_{1(1),\q}\nonumber\\&-18 W_{(1)}^2 C_{1(2),\q}+6 \tilde{W}_{(4),u}+12 \b_{(3)} W_{(1),u}+24 \b_{(2)} W_{(2),u}-12 \tilde{W}_{(4)} \beta_{(0),u}\nonumber\\&+48
	W_{(4)} \beta_{(0),u}-48 W_{(2)} \b_{(2)} \beta_{(0),u}-24 W_{(1)}\b_{(3)} \beta_{(0),u}+24 W_{(2)} \b_{(2),u}\nonumber\\&+12 W_{(1)} \b_{(3),u}-12 W_{(3)} C_{1(1),u}-15 W_{\left(\frac{5}{2}\right)}
	C_{1\left(\frac{3}{2}\right),u}-18 W_{(2)} C_{1(2),u}\nonumber\\&-24 W_{(1)} C_{1(3),u}-15 W_{\left(\frac{5}{2}\right),u} C_{1\left(\frac{3}{2}\right)}+30 W_{\left(\frac{5}{2}\right)} \beta_{(0),u}
	C_{1\left(\frac{3}{2}\right)}-6 W_{(1)} C_{1\left(\frac{3}{2}\right),u} C_{1\left(\frac{3}{2}\right)}\nonumber\\&-18 U_{(0)} W_{(1)} C_{1(2)}+6 W_{(1)} W_{(1),\q} C_{1(2)}+12 W_{(1)}^2 \beta_{(0),\q} C_{1(2)}-12 W_{(2),u} C_{1(2)}\nonumber\\&+24 W_{(2)} \beta_{(0),\q} C_{1(2)}-6 W_{(1),u} C_{1(3)}+12 W_{(1)} \beta_{(0),u} C_{1(3)}-6 W_{(2)}^2 C_{1(0),\q}\nonumber\\&-12 W_{(1)} W_{(3)} C_{1(0),\q}-12 W_{(1)}^2 \b_{(2)} C_{1(0),\q}+6 W_{(1)}^2 C_{1(2)} C_{1(0),\q}
	\end{align}	
\endgroup

	\section{Iterative differentiation approach} \label{appB}

	In analysing equations asymptotically as $r\rightarrow \infty$, it is more elegant to change the radial variable so that the boundary is at $z \rightarrow 0$. One may then proceed to determine the asymptotic solution via an iterative differentiation procedure as in AdS \cite{deHaro2001,Poole:2018koa}. 
	
	To illustrate this, let us implement the change of variables $r = 1/z^2$ in the first three main equations \eqref{Rrr},\eqref{Rrt} and \eqref{Rtrace}
	\be
	\b_{,z}=-\frac{z}{24}\left((C_{1,z})^2+(C_{2,z})^2+(C_{3,z})^2\right) \label{b1}
	\ee
	\begin{align}
	\frac{z^9}{4}\partial_z\left(\frac{1}{2z^7}e^{C_1-2\b}W_{,z}\right)&= \frac{ze^{C_1-2\b}}{4}\left(zW_{,zz}+zC_{1,z}W_{,z}-2z\b_{,z}W_{,z}-7W_{,z}\right) \nonumber \\& -\frac{z^2}{2}\left(6\b_{,\q}+z\b_{,\q z}\right)+\frac{z^3}{4}\left((\cot\q-2\tan\q)C_{1,z}+C_{1,z\q}\right) \label{b2} \\&
	-\frac{z^3C_{1,z}}{8}(2C_{1,\q}+C_{2,\q})-\frac{z^3 C_{2,z}}{8}\left(C_{1,\q}+2C_{2,\q}+\frac{2}{\sin\q\cos\q}\right) \nonumber
	\end{align}
	
	\begin{align}
	3z^4(2-z\partial_z)\mathcal{U}&=-\frac{z^4e^{2\b-C_1}}{2}\sec\q\csc\q( C_{1,\q}+2C_{2,\q}) \label{b3} \\&
	+z^4e^{-C_1}(\cot\q-\tan\q)\left(\frac{e^{C_1}}{2}(12z^{-2}-z^{-1}\partial_z)W-2 e^{2\beta}\beta_{,\q}+\frac{5}{2}e^{2\beta} C_{1,\q}\right)\nonumber\\&
	+\frac{z^4e^{2\b-C_1}}{2}\left(12-(2\b_{,\q})^2+4\beta_{,\q} C_{1,\q}-2( C_{1,\q})^2- C_{1,\q} C_{2,\q}-( C_{2,\q})^2\right)\nonumber\\&+\frac{z^4e^{2\b-C_1}}{2}\left(-4\beta_{,\q\q}+2C_{1,\q\q}\right)	-\frac{z^2}{8}e^{-4\b+C_1}(W_{,z})^2\nonumber\\&
	-3z^{2}\left(2-\frac{z}{6}\partial_z\right) W_{,\q} \nonumber
	\end{align}
	The reason for the specific choice of variable $z$ is that the powers in the resulting asymptotic series will then be integer.
		
Let us now explain the iterative differentiation approach, beginning with equation \eqref{b1}. Taking the limit of \eqref{b1} as $z \rightarrow 0$ we obtain
\be
[ \b_{,z} ]_{z=0} = 0
\ee	
provided that $z^{\frac{1}{2}} C_{i,z} \rightarrow 0$ as $z \rightarrow 0$; this limit can be justified using the last of the main equations, applying the given boundary conditions. Clearly \eqref{b1} does not impose any restrictions on $[\beta]_{z=0}$, while the equation above implies that the term in the asymptotic expansion of $\beta$ at order $z$ vanishes, as we found before. Differentiating \eqref{b1} and taking the limit as $z \rightarrow 0$ then gives
\be
[ \b_{,zz} ]_{z=0} = -\frac{1 }{24} \left [ (C_{1,z})^2+(C_{2,z})^2+(C_{3,z})^2\right ]_{z=0} 
\ee	
Continuing the process, we will clearly determine the derivatives of $\beta$ as $z \rightarrow 0$ to all orders. 

Just as in the standard Bondi analysis, we solve the differential equations in the nested order, substituting the solution of \eqref{b1} into the righthandside of \eqref{b2}, and then using the solutions of both \eqref{b1} and \eqref{b2} in \eqref{b3}. To understand how logarithms arise in the asymptotic expansion it is useful to rewrite \eqref{b3} in the form
\be
(2 - z \partial_z ) {\cal U} = {\cal P} [\beta,C_i,W], \label{b6}
\ee
where ${\cal P}$ is implicitly a functional of the functions $(\beta,C_i,W)$ and their derivatives. Using the solutions of the other main equations one can show that 
\be
[{\cal P}]_{z=0} \qquad
[{\cal P}_{,z}]_{z=0} \qquad 
[ {\cal P}_{,zz}]_{z=0}
\ee
are all non-vanishing for generic boundary data $([\beta]_{z=0}, [C_i]_{z=0})$. From \eqref{b6} and its first derivative one obtains
\be
[ {\cal U} ]_{z=0} = \frac{1}{2} [{\cal P}]_{z=0} \qquad
[ {\cal U}_{,z} ]_{z=0}  =  [{\cal P}_{,z}]_{z=0} 
\ee
but differentiating again one obtains 
\be
[ z \partial_{z}^3 {\cal U}]_{z=0} = -  [ {\cal P}_{,zz}]_{z=0}
\ee
i.e. $ [ {\cal U}_{,zz} ]_{z=0}$ is unconstrained, while $\partial_{z}^3 {\cal U}$ has a first order pole at $z \rightarrow 0$. The latter induces the logarithmic term in the asymptotic expansion at order $z^2$.

	\addcontentsline{toc}{section}{Bibliography}
	\bibliography{refs}
	
	
	

\end{document}